\documentclass[aps,prd,nofootinbib,preprintnumbers, twocolumn,amssymb,eqsecnum,notitlepage]{revtex4-1}

\usepackage[utf8]{inputenc}  
\usepackage[T1]{fontenc}     
\usepackage{bm}
\usepackage{amsmath,amssymb,lmodern} 
\usepackage{sidecap}
\usepackage[caption=false]{subfig}
\usepackage{graphicx}
\usepackage{tabularx}
\usepackage{enumerate} 
\usepackage{makecell}
\usepackage{hyperref}\hypersetup{
    colorlinks,%
    citecolor=blue,%
    filecolor=blue,%
    linkcolor=blue,%
    urlcolor=blue
}
\usepackage{color}
\usepackage{cleveref}
\allowdisplaybreaks[1]

\newcommand{\newc}{\newcommand}
\newc{\tif}{\tilde{f}}
\newc{\tih}{\tilde{h}}
\newc{\tip}{\tilde{\phi}}
\newc{\tiA}{\tilde{A}}

\newcommand{\mr}[1]{\mathrm{#1}}
\newcommand{\mL}[1]{\mathcal{#1}}
\newcommand{\itl}[1]{\textit{#1}}

\newcommand{\ben}{\begin{eqnarray}}
\newcommand{\een}{\end{eqnarray}}
\newc{\be}{\begin{equation}}
\newc{\ee}{\end{equation}}
\newc{\ba}{\begin{eqnarray}}
\newc{\ea}{\end{eqnarray}}
\newc{\bea}{\begin{eqnarray*}}
\newc{\eea}{\end{eqnarray*}}
\newc{\D}{\partial}
\newc{\ie}{{\it i.e.} }
\newc{\eg}{{\it e.g.} }
\newc{\etc}{{\it etc.} }
\newc{\etal}{{\it et al.}}

\newc{\ra}{\rightarrow}
\newc{\lra}{\leftrightarrow}
\newc{\lsim}{\buildrel{<}\over{\sim}}
\newc{\gsim}{\buildrel{>}\over{\sim}}
\newc{\aP}{\alpha_{\rm P}}
\newc{\dphi}{\delta\phi}
\newc{\da}{\delta A}
\newc{\tp}{\dot{\phi}}
\newc{\Ve}{V_{\rm eff}}
\newc{\Vep}{V_{{\rm eff},\phi}}

\begin{document}
\preprint{PI/UAN-2019-647FT}

\title{Anisotropic \texorpdfstring{$\boldsymbol{2}$}{2}-form dark energy}

\author{Juan P.~Beltr\'an Almeida$^{1}$, 
Alejandro Guarnizo$^{1,2}$, Ryotaro Kase$^{3}$, 
Shinji Tsujikawa$^{3}$, and 
C\'esar A.~Valenzuela-Toledo$^{2}$}

\affiliation{
$^1$Departamento de F\'isica, Facultad de Ciencias, Universidad Antonio Nari\~no, \\ Cra 3 Este \# 47A-15, 
Bogot\'a DC, Colombia\\
$^2$Departamento de F\'isica, Universidad del Valle,\\
Ciudad Universitaria Mel\'endez, Santiago de Cali 760032, Colombia\\
$^3$Department of Physics, Faculty of Science, Tokyo University of Science, 1-3, Kagurazaka,
Shinjuku-ku, Tokyo 162-8601, Japan\\
}

\begin{abstract}

We study the dynamics of dark energy in the presence of a 2-form field 
coupled to a canonical scalar field $\phi$. 
We consider the coupling proportional to 
$e^{-\mu \phi/M_{\rm pl}} H_{\alpha \beta \gamma}H^{\alpha \beta \gamma}$ 
and the scalar potential $V(\phi) \propto e^{-\lambda \phi/M_{\rm pl}}$, 
where $H_{\alpha \beta \gamma}$ 
is the 2-form field strength, $\mu, \lambda$ are constants, and $M_{\rm pl}$ 
is the reduced Planck mass. 
We show the existence of an anisotropic matter-dominated scaling solution 
followed by a stable accelerated fixed point with a non-vanishing shear.
Even if $\lambda \geq {\cal O}(1)$, it is possible to realize the 
dark energy equation of state $w_{\rm DE}$ close to $-1$ 
at low redshifts for $\mu \gg \lambda$. 
The existence of anisotropic hair and the oscillating behavior of 
$w_{\rm DE}$ are key features for 
distinguishing our scenario from other dark energy 
models like quintessence.

\end{abstract}


\maketitle

%%%%%%%%%%%%%%%%%%%%%%%%
\section{Introduction}
%%%%%%%%%%%%%%%%%%%%%%%%

Since the first discovery of late-time cosmic acceleration 
from the distant supernovae type Ia (SN Ia) \cite{SN1,SN2}, 
the origin of this phenomenon has not been identified yet. 
The cosmological constant is a simplest 
candidate for dark energy, but if it originates from 
vacuum energy associated with particle physics, 
it is plagued by a huge 
energy gap between its observed value and the 
theoretically predicted value \cite{Weinberg}. 
Instead, there are dynamical dark energy 
models dubbed quintessence in which a canonical 
scalar field $\phi$ slowly evolving along a potential 
$V(\phi)$ leads to a time-varying field equation 
of state \cite{quin1,quin2,quin3,quin4,quin5,quin6,quin7}.

In quintessence, the condition for cosmic acceleration 
can be quantified by the dimensionless parameter $\lambda=-M_{\rm pl}V_{,\phi}/V$, 
where $M_{\rm pl}$ is the reduced Planck mass and 
$V_{,\phi}={\rm d}V/{\rm d}\phi$.
For constant $\lambda$, i.e., for the exponential potential 
$V(\phi)=V_0 e^{-\lambda \phi/M_{\rm pl}}$, the accelerated expansion occurs 
for $|\lambda|<\sqrt{2}$ \cite{CLW,CST,Tsuji13}. 
Under this condition, the solutions finally approach an 
attractor characterized by the dark energy equation of 
state $w_{\rm DE}=-1+\lambda^2/3$.

In the context of higher-dimensional theories like string/M 
theories, the exponential potential can arise from compactifications 
in hyperbolic manifolds or S-brane solutions \cite{Garriga:2000cv,Emparan:2003gg}. 
After the dimensional reduction, the slope 
$|\lambda|$ is typically larger than the order 1. 
In this case the accelerated attractor mentioned 
above is not present, while the temporal cosmic 
acceleration is possible for the internal manifold 
changing in time \cite{Townsend:2003fx,Ohta:2003pu,Wohlfarth:2003ni,Roy:2003nd}.
The construction of a meta-stable de Sitter vacuum 
in string theory also suggested the swampland conjecture
stating that $|\lambda|$ has a lower bound of order 1 \cite{swamp1,swamp2}. 
It is worthy of pursuing possibilities for realizing 
the cosmic acceleration even for steep scalar potentials 
satisfying $|\lambda|\ge {\cal O}(1)$. 

In string theory, there are $p$-form fields arising from the 
Ramond-Ramond sector \cite{string}. 
The 1-form field, which corresponds to a vector field $A_{\mu}$, 
can be generally coupled to 
a scalar (0-form) field $\phi$ \cite{Almeida:2018fwe}. 
The commonly studied coupling in the cosmological 
context has the form 
$-f_1(\phi)F_{\mu \nu}F^{\mu \nu}/4$, where 
$f_1(\phi)$ is a function of $\phi$ and 
$F_{\mu \nu}=\partial_{\mu} A_{\nu}
-\partial_{\nu} A_{\mu}$ is the field strength 
tensor \cite{Ratra1991bn,Bamba2003av,Martin2007ue,Yokoyama2008xw,Dimopoulos2009am,Dimopoulos2009vu,Fujita:2018zbr}. 
During the inflationary period, it is known that the 
vector field 
can generate the non-vanishing anisotropic shear for 
a suitable choice of the coupling $f_1(\phi)$ related to 
the scalar potential 
$V(\phi)$ \cite{Watanabe,Watanabe2,Kanno:2010nr,Ohashi:2013pca}. 
The anisotropic hair sustained during inflation can
leave several interesting observational signatures for the 2-point and 3-point correlation functions of 
Cosmic Microwave Background (CMB) 
temperature anisotropies \cite{Gumrukcuoglu2010yc,Gum,Himmetoglu2009mk,Bartolo,Namba2012gg,Shiraishi2013vja,Fujita2013pgp}. 

The 2-form field $B_{\alpha \beta}$ coupled to 
the scalar  field $\phi$ through the form 
$-f_2(\phi)H_{\alpha \beta \gamma}H^{\alpha \beta \gamma}/12$, 
where $f_2(\phi)$ is a function of $\phi$ and 
$H_{\alpha \beta \gamma}$ is the field strength of 
$B_{\alpha \beta}$, can also give rise to anisotropic inflation 
for an appropriate choice of 
$f_2(\phi)$ \cite{Ohashi2,Ito,Obata:2018ilf,Almeida:2019xzt}. 
The observational signatures in CMB imprinted by 
the 2-form is different from those by the 1-form, so they can be distinguished between each other from the 
scalar power spectrum and primordial 
non-Gaussianities \cite{Ohashi:2013qba}.
If the anisotropic shear does not survive either 
during inflation or in the later cosmological epoch, 
the 2-form energy density 
decreases as $\rho_B \propto a^{-2}$, where 
$a$ is the isotropic scale factor. 
This is in contrast to the 1-form field,  whose energy density 
decreases as radiation ($\rho_A \propto a^{-4}$) 
in the isotropic context.
Hence the energy density of 2-form can be 
generally prominent at late times compared to that of 1-form. 

For the 1-form field coupled to a dark energy field $\phi$, 
Thorsrud {\it et al.} \cite{Thorsrud:2012mu} studied the late-time cosmological 
dynamics in the presence of an additional coupling between 
$\phi$ and matter. They found interesting anisotropic scaling solutions relevant to the matter and 
dark energy dominated epochs. 
The existence of non-vanishing anisotropic shear after the decoupling epoch leaves modifications to the observables in CMB 
and SN Ia measurements \cite{Koivisto:2007bp,Koivisto:2008ig,Battye:2009ze,Appleby:2009za,Campanelli:2010zx,Appleby:2012as}.

In this paper, we study the late-time cosmology in the presence of the interaction 
$-f(\phi)H_{\alpha \beta \gamma}H^{\alpha \beta \gamma}/12$ between the 2-form and 
the scalar field $\phi$. 
We consider the exponential potential 
$V(\phi)=V_0 e^{-\lambda \phi/M_{\rm pl}}$ 
for the scalar sector and adopt the coupling 
of the form $f(\phi)=f_0 e^{-\mu \phi/M_{\rm pl}}$.
We show that, even for $\lambda \ge {\cal O}(1)$, the 
late-time cosmic acceleration with the dark energy 
equation of state  $w_{\rm DE}$ close to $-1$ can be 
realized for the coupling constant $\mu$ in the range 
$\mu \gg \lambda$. Thus, this model is an explicit 
example where the accelerated expansion consistent 
with current observations \cite{Betoule:2014frx,Aghanim:2018eyx} 
is possible even with a 
steep exponential potential.  

Moreover, we show that the radiation-dominated epoch 
with an initially negligible anisotropic shear 
is followed by the scaling matter era with a non-vanishing 
anisotropic hair. As long as the anisotropic dark energy 
dominated fixed point is present, it is a stable spiral 
for $\mu \gg \lambda \ge {\cal O}(1)$. 
Thus, our model gives rise to several interesting observational signatures such as the surviving anisotropic 
shear after the radiation era 
and the oscillating dark energy equation of state in 
the range $w_{\rm DE}>-1$.

This paper is organized as follows. 
In Sec.~\ref{sec:syspf}, we derive the background equations of motion 
in the presence of a perfect fluid 
on the anisotropic cosmological background.
In Sec.~\ref{sec:dyna}, we obtain the fixed points 
associated with radiation, matter, and dark energy 
dominated epochs and discuss the stabilities of them.
In Sec.~\ref{sec:cossec}, we study the cosmological dynamics 
in our model by paying particular attention to the evolution 
of $w_{\rm DE}$ and the anisotropic shear. 
Finally, Sec.~\ref{sec:consec} is devoted to conclusions. 
Throughout the paper, we use the Lorentzian metric $g_{\mu\nu}$ with the sign 
convention $(-,+,+,+)$, and greek indices as
$\alpha, \beta, \gamma \cdots$, will denote space-time coordinates. 

%%%%%%%%%%%%%%%%%%%%%%%%%%%%%
\section{Background equations of motion}
\label{sec:syspf}
%%%%%%%%%%%%%%%%%%%%%%%%%%%%%

In the 4-dimensional space-time, we consider a 2-form field 
$B_{\alpha \beta}$ with the field strength 
$H_{\alpha \beta \gamma}=
3 \partial_{[\alpha}B_{\beta \gamma]}$. 
We also take into account a canonical scalar field 
$\phi$ with the potential $V(\phi)$ and assume that 
$\phi$ is coupled to the 2-form through the interacting 
Lagrangian $-f(\phi) H_{\alpha \beta \gamma}
H^{\alpha \beta \gamma}/12$.
We do not consider non-minimal couplings to gravity, 
so the gravity sector is described 
by the  Einstein-Hilbert Lagrangian $M_{\rm pl}^2 R/2$, 
where $R$ is the Ricci scalar. We also add
a perfect fluid described by the purely k-essence 
Lagrangian $P_f(Z)$, where 
$Z=-\partial_{\mu} \chi \partial^{\mu} \chi/2$ is 
the kinetic term of a scalar
field $\chi$ \cite{Hu05,Arroja,KT14}.
Then, the action of our theory is given by 
\ba
\hspace{-0.7cm}
\mL{S} 
&=& \int \mr{d}^4 x \sqrt{-g}  
\biggl[  \frac{M_{\rm{pl}}^2}{2}R
-\frac{1}{2} \partial_{\mu}\phi  \partial^{\mu}\phi
-V(\phi) \nonumber \\
\hspace{-0.7cm}
& &\qquad \qquad \qquad
-\frac{1}{12} f(\phi) H_{\alpha \beta \gamma}
H^{\alpha \beta \gamma} 
+P_f(Z) \biggr]\,,
\label{eq:LT}
\ea
where $g$ is the determinant of metric tensor 
$g_{\mu \nu}$ and $f(\phi)$ is a function of $\phi$.

Let us derive the dynamical equations of motion for the action 
\eqref{eq:LT} on the anisotropic cosmological background. 
We consider the configuration in which the 2-form field 
is in the $(y,z)$ plane, such that 
\be
B_{\alpha \beta} \mr{d}x^{\alpha} 
\wedge \mr{d}x^{\beta}
=2v_B(t) \mr{d}y \wedge \mr{d}z\,,
\ee
where $v_B$ depends on the cosmic time $t$.
In the $(y,z)$ plane there is a rotational symmetry, so  
the line element can be taken as 
\be
\mr{d}s^2 = -N(t)^2 \mr{d}t^2 
+ e^{2\alpha(t)} \left[ e^{-4\sigma (t)}\mr{d}x^2
+e^{2\sigma (t)}(\mr{d}y^2+\mr{d}z^2) \right],
\label{anisotropic-metric}
\ee
where $N(t)$ is the lapse function, $a \equiv e^{\alpha(t)}$ is 
the geometric mean of three scale factors 
(with the normalization $a=1$ today), and $\sigma(t)$ 
is the spatial shear. 
The non-vanishing components of $B_{\alpha \beta}$ 
are $B_{23}=-B_{32}=v_B$. 
On the background (\ref{anisotropic-metric}), the action 
\eqref{eq:LT} is expressed as
\ba
{\cal S}
&=&
\int {\rm d}^4 x \biggl[ \frac{3M_{\rm pl}^2 
e^{3\alpha}}{N} \left(\dot{\sigma}^2 -\dot{\alpha}^2 
\right)+e^{3\alpha} \left\{ \frac{\dot{\phi}^2}{2N}
-NV(\phi) \right\} \nonumber \\
& & \qquad \quad
+\frac{f(\phi)}{2N}e^{-\alpha-4\sigma} \dot{v}_B^2
+ Ne^{3\alpha} P_f(Z(N)) \biggr],
\label{ac2}
\ea
where $Z=\dot{\chi}^2/(2N^2)$ and 
a dot represents a derivative with respect to $t$. 

Varying the action (\ref{ac2}) with respect to 
$N, \alpha, \sigma, \phi$, {\bf $\chi$} and setting $N=1$ 
at the end, we obtain
\ba
\hspace{-0.5cm}
& &
3M_{\rm pl}^2 H^2 \left( 1-\Sigma^2 \right)
=\frac{1}{2} \dot{\phi}^2+V(\phi)+\rho_B+\rho_f\,,
\label{eq:be1} \\
\hspace{-0.5cm}
& &
M_{\rm pl}^2 \left( \dot{H}+3H^2 \Sigma^2 \right)
=-\frac{1}{2} \dot{\phi}^2 
-\frac{1}{3} \rho_B
-\frac{1}{2} \left( \rho_f+P_f \right),
\label{eq:be2} \\
\hspace{-0.5cm}
& &
M_{\rm pl}^2 \left[ H\dot{\Sigma}
+\left( \dot{H}+3H^2 \right) \Sigma \right]
=-\frac{2}{3} \rho_B\,,
\label{eq:be3} \\
\hspace{-0.5cm}
& &
\ddot{\phi}+3H \dot{\phi}+V_{,\phi}
-\frac{f_{,\phi}}{f} \rho_B
=0\,,\label{eq:be4} \\
\hspace{-0.5cm}
& &
\dot{\rho}_f+3H \left( \rho_f+P_f \right)=0\,,
\label{eq:be5}
\ea
being
\be
H \equiv \dot{\alpha}\,,\qquad 
\Sigma \equiv \frac{\dot{\sigma}}{H}\,,
\ee
and $\rho_B$ and $\rho_f$ correspond to the energy densities 
of the 2-form and the perfect fluid, defined, respectively, by 
\be
\rho_B=\frac{f(\phi)}{2} 
e^{-4\alpha-4\sigma} \dot{v}_B^2\,,\qquad 
\rho_f=\dot{\chi}^2 P_{f,Z}-P_f\,.
\label{rhoAB}
\ee
Note that we used the notation in which a comma in 
the subscript represents a derivative with respect to a 
corresponding variable, e.g., 
$f_{,\phi}={\rm d} f/{\rm d}  \phi$.
Varying the action (\ref{ac2}) with respect to $v_B$, 
it follows that 
\be
f(\phi) e^{-\alpha - 4 \sigma}\dot{v}_B
=p_B\,, 
\label{eq:be6} 
\ee
where $p_B$ is a constant.  
Taking the time derivative of $\rho_B$ in 
Eq.~(\ref{rhoAB}) and using Eq.~(\ref{eq:be6}), 
the 2-form energy density $\rho_B$ obeys 
the differential equation, 
\be
\dot{\rho}_B=-2H \rho_B \left( 1-2\Sigma
+ \frac{\dot{f}}{2H f} 
 \right)\,.
\label{drhoB}
\ee

For the scalar potential $V(\phi)$ and the coupling $f(\phi)$, 
we adopt the exponential functions 
given by \cite{Ito,Kanno:2010nr,Ohashi:2013pca}:
\be
V(\phi)=V_0 e^{-\lambda \phi/M_{\rm pl}}\,,\qquad 
f(\phi)=f_0 e^{-\mu \phi/M_{\rm pl}}\,,
\label{f12}
\ee
where $V_0, \lambda, f_0, \mu$ are assumed to be 
positive constants.  
We are interested in the case in which the late-time 
cosmic acceleration can be realized for 
\be
\lambda \ge {\cal O}(1)\,,\qquad \mu \ge {\cal O}(1)\,,
\label{lammu}
\ee
whose conditions are assumed in the following.
If the coupling $f(\phi)$ is absent, 
the cosmic acceleration with the dark energy 
equation of state close to $-1$ occurs only for 
$\lambda^2 \ll 2$ \cite{CLW,CST,Tsuji13}.

%%%%%%%%%%%%%%%%%%%%%%%%%%%%%
\section{Dynamical system and fixed points}
\label{sec:dyna}
%%%%%%%%%%%%%%%%%%%%%%%%%%%%%

We express the background equations of motion derived in 
Sec.~\ref{sec:syspf} in an autonomous form and obtain 
the corresponding fixed points.
For the perfect fluid given by the Lagrangian $P_f (Z)$, 
we take into account both 
non-relativistic matter (energy density $\rho_m$ and
negligible pressure) and radiation 
(energy density $\rho_r$ and pressure $P_r=\rho_r/3$), 
so that $\rho_f=\rho_m+\rho_r$ and $P_f=\rho_r/3$.

\subsection{Dynamical system}

We introduce the following dimensionless quantities:
\ba
& &
x_1=\frac{\dot{\phi}}{\sqrt{6}H M_{\rm pl}}\,,\quad 
x_2=\frac{\sqrt{V}}{\sqrt{3}H M_{\rm pl}}\,,\quad 
\Omega_B=\frac{\rho_B}{3H^2 M_{\rm pl}^2}\,,
\nonumber \\
& & 
\Omega_r=\frac{\rho_r}{3H^2 M_{\rm pl}^2}\,,\quad 
\Omega_m=\frac{\rho_m}{3H^2 M_{\rm pl}^2}\,.
\ea
{}From Eq.~(\ref{eq:be1}), there is the constraint
\be
\Omega_m=1-x_1^2-x_2^2-\Sigma^2
-\Omega_B-\Omega_r\,.
\label{coneq}
\ee
By using Eqs.~(\ref{eq:be2}) and (\ref{eq:be4}) 
with Eq.~(\ref{coneq}), we obtain
\ba
\frac{\dot{H}}{H^2} &=& 
-\frac{1}{2} \left( 3+3x_1^2-3x_2^2+3\Sigma^2 
-\Omega_B+\Omega_r \right)\,,
\label{eq:be2d} \\
\frac{\ddot{\phi}}{H \dot{\phi}} &=&
-3+\frac{\sqrt{6}}{2x_1} \left( \lambda x_2^2 
-\mu  \Omega_B \right)\,.
\label{eq:be4d}
\ea

{}From Eqs.~(\ref{eq:be3}), (\ref{eq:be5}), (\ref{coneq}), 
(\ref{eq:be2d}), and (\ref{eq:be4d}), the dimensionless variables 
$x_1, x_2, \Sigma, \Omega_B$, and $\Omega_r$ obey
\ba
\hspace{-0.5cm}
x_1' &=& \frac{3}{2}x_1 \left( x_1^2-x_2^2+\Sigma^2 -1
-\frac{1}{3} \Omega_B+\frac{1}{3} \Omega_r
\right) \nonumber \\
\hspace{-0.5cm}
& &+\frac{\sqrt{6}}{2} \left( \lambda x_2^2 -\mu \Omega_B 
\right) \,, \label{auto1}\\
\hspace{-0.5cm}
x_2' &=& \frac{1}{2} x_2 ( 3 x_1^2-3x_2^2+3\Sigma^2+3
-\sqrt{6} \lambda x_1 \nonumber \label{auto2} \\
\hspace{-0.5cm}
& &
-\Omega_B+\Omega_r )\,,\\
\hspace{-0.5cm}
\Sigma' &=& \frac{1}{2} \Sigma \left( 3x_1^2
-3x_2^2+3\Sigma^2-3-\Omega_B+\Omega_r \right) 
\nonumber \label{auto3} \\
\hspace{-0.5cm}
& &
-2\Omega_B\,,\\
\hspace{-0.5cm}
\Omega_B' &=& \Omega_B ( 3x_1^2
-3x_2^2+3 \Sigma^2+4\Sigma+1+\sqrt{6} \mu x_1 
\nonumber \label{auto4} \\
\hspace{-0.5cm}
& &-\Omega_B+\Omega_r )\,,\\
\Omega_r' &=& \Omega_r \left( 3x_1^2-3x_2^2 
+3\Sigma^2-1-\Omega_B+\Omega_r \right)\,,\label{auto5}
\ea
where a prime represents a derivative with respect to the 
number of e-foldings $\alpha=\ln a$. 
The cosmological dynamics is known by solving 
Eqs.~(\ref{auto1})-(\ref{auto5}) with Eq.~(\ref{coneq})
for given initial values of $x_1, x_2, \Sigma, \Omega_B, 
\Omega_r$.

The effective equation of state, which is defined by 
$w_{\rm eff} \equiv -1-2\dot{H}/(3H^2)$, 
characterizes the evolution of mean scale factor $a(t)$.
{}From Eq.~(\ref{eq:be2d}), it follows that 
\be
w_{\rm eff}=x_1^2-x_2^2+\Sigma^2
-\frac{1}{3}\Omega_B+\frac{1}{3}\Omega_r\,.
\ee
The radiation- and matter-dominated epochs correspond to 
$w_{\rm eff} \simeq 1/3$ and $w_{\rm eff} \simeq 0$, 
respectively. The cosmic acceleration occurs 
for $w_{\rm eff}<-1/3$.
We can express Eqs.~(\ref{eq:be1}) and (\ref{eq:be2})
in the form:
\ba
& &
3M_{\rm pl}^2 H^2=\rho_{\rm DE}+\rho_r+\rho_m\,,\\
& &
2M_{\rm pl}^2 \dot{H}=-\rho_{\rm DE}-P_{\rm DE}
-\frac{4}{3}\rho_r 
-\rho_m\,,
\ea
where 
\ba
\rho_{\rm DE}
&=&
\frac{1}{2} \dot{\phi}^2+V(\phi)+\rho_B
+3M_{\rm pl}^2 H^2 \Sigma^2\,,
\label{rhoDE}\\
P_{\rm DE}
&=&
\frac{1}{2} \dot{\phi}^2-V(\phi)
-\frac{1}{3} \rho_{B}
+3M_{\rm pl}^2 H^2 \Sigma^2\,.
\label{PDE}
\ea
Defining the density parameter and the equation of state arising from 
the dark sector, 
as $\Omega_{\rm DE}=\rho_{\rm DE}/(3H^2 M_{\rm pl}^2)$ 
and $w_{\rm DE}=P_{\rm DE}/\rho_{\rm DE}$, respectively,
it follows that 
\ba
\Omega_{\rm DE} &=& x_1^2+x_2^2+\Sigma^2+\Omega_B
=1-\Omega_r-\Omega_m\,,\label{Omede} \\
w_{\rm DE} &=& \frac{3(x_1^2-x_2^2+\Sigma^2)-\Omega_B}
{3(x_1^2+x_2^2+\Sigma^2+\Omega_B)}\,,
\label{wde}
\ea
where we used Eq.~(\ref{coneq}) in the second 
equality of Eq.~(\ref{Omede}). 
The above definitions of $\Omega_{\rm DE}$ and $w_{\rm DE}$ are 
not the same as those given in Refs.~\cite{Thorsrud:2012mu,Appleby:2009za,Campanelli:2010zx}, 
because, in our case, the right hand sides of Eqs.~(\ref{rhoDE}) and (\ref{PDE}) 
contain the spatial shear terms $3M_{\rm pl}^2 H^2 \Sigma^2$.
As we will see later in Sec.~\ref{sec:cossec}, the CMB and SN Ia
data give the bound $|\Sigma| \ll 1$, which limits the model parameter 
space in the range $\mu \gg \lambda \geq {\cal O}(1)$.
In such cases, the values of $\Omega_{\rm DE}$ and $w_{\rm DE}$ 
computed from Eqs.~(\ref{Omede}) and (\ref{wde}) are 
similar to those evaluated without the spatial shear terms $\Sigma^2$.

\subsection{Fixed points}
\label{fixedsec}

The fixed points of the dynamical system can be derived by setting 
$x_1'=0, x_2'=0, \Sigma'=0, \Omega_B'=0, 
\Omega_r'=0$ in Eqs.~(\ref{auto1})-(\ref{auto5}) 
and solving the corresponding algebraic equations. 
In what follows, we show the fixed points relevant to 
the radiation era ($\Omega_r \simeq 1$, $w_{\rm eff} \simeq 1/3$), 
matter era ($\Omega_m \simeq 1$, $w_{\rm eff} \simeq 0$), 
and accelerated epoch ($\Omega_{\rm DE} \simeq 1$, 
$w_{\rm eff}<-1/3$).
There are two additional points (\itl{a4}) and (\itl{b4}) 
presented in Appendix~\ref{a4b4}. 
For $\lambda$ and $\mu$ in the range (\ref{lammu}), however, 
they are irrelevant to the realistic cosmological sequence.

\subsubsection{Radiation dominance}

$\bullet$ (\itl{a1}) Isotropic radiation point
\vspace{0.1cm}
\ba
& &
x_1=0\,,\quad x_2=0\,,\quad \Sigma=0\,,\nonumber \\
& &
\Omega_B=0\,,\quad \Omega_r=1\,,\quad \Omega_m=0\,,
\ea
with $\Omega_{\rm DE}=0$ and $w_{\rm DE}$ undetermined.

\vspace{0.2cm}

$\bullet$ (\itl{a2}) Isotropic radiation scaling solution 
\vspace{0.1cm}
\ba
& &
x_1=\frac{2\sqrt{6}}{3\lambda}\,,\quad 
x_2=\frac{2\sqrt{3}}{3\lambda}\,,\quad 
\Sigma=0\,,\nonumber \\
& &
\Omega_B=0\,,\quad \Omega_r=1-\frac{4}{\lambda^2}\,,
\quad \Omega_m=0\,,
\ea
with $\Omega_{\rm DE}=4/\lambda^2$ and $w_{\rm DE}=1/3$.
The energy density of dark energy scales in the same 
manner as that of radiation. 
The big-bang nucleosynthesis (BBN) constraint gives the bound 
$\Omega_{\rm DE}<0.045$ \cite{Bean}, which 
translates to $\lambda>9.4$ \cite{Ohashi:2009xw}.

\vspace{0.2cm}

$\bullet$ (\itl{a3}) Anisotropic radiation point
\vspace{0.1cm}
\ba
& &
x_1=-\frac{\sqrt{6}\mu}{3\mu^2+8}\,,\quad 
x_2=0\,,\quad \Sigma=-\frac{4}{3\mu^2+8}\,,\nonumber \\
& &
\Omega_B=\frac{2}{3\mu^2+8}\,,\quad 
\Omega_r=\frac{3\mu^2+4}{3\mu^2+8}\,,
\quad \Omega_m=0\,,
\label{a3}
\ea
with $\Omega_{\rm DE}=4/(3\mu^2+8)$ and $w_{\rm DE}=1/3$.
This is a scaling solution with the non-vanishing anisotropic 
shear ($\Sigma \neq 0$). 
The BBN constraint $\Omega_{\rm DE}<0.045$
gives the bound {\bf $\mu>5.2$}.

\subsubsection{Matter dominance}

$\bullet$ (\itl{b1}) Isotropic matter point
\vspace{0.1cm}
\ba
& &
x_1=0\,,\quad x_2=0\,,\quad \Sigma=0\,,\nonumber \\
& &
\Omega_B=0\,,\quad \Omega_r=0\,,\quad \Omega_m=1\,,
\ea
with $\Omega_{\rm DE}=0$ and $w_{\rm DE}$ undetermined.

\vspace{0.2cm}

$\bullet$ (\itl{b2}) Isotropic matter scaling solution 
\vspace{0.1cm}
\ba
& &
x_1=\frac{\sqrt{6}}{2\lambda}\,,\quad 
x_2=\frac{\sqrt{6}}{2\lambda}\,,\quad \Sigma=0\,,\nonumber \\
& &
\Omega_B=0\,,\quad \Omega_r=0\,,
\quad \Omega_m=1-\frac{3}{\lambda^2}\,,
\ea
with $\Omega_{\rm DE}=3/\lambda^2$ and $w_{\rm DE}=0$. 
{}From the Planck CMB data, the dark energy density parameter 
is constrained to be $\Omega_{\rm DE}<0.02$ around 
the redshift $50$ \cite{Ade15}, 
which translates to $\lambda>12$. 

\vspace{0.2cm}

$\bullet$ (\itl{b3}) Anisotropic matter point 
\vspace{0.1cm}
\ba
\hspace{-0.7cm}
& &
x_1=-\frac{\sqrt{6}\mu}{2(3\mu^2+8)}\,,\quad 
x_2=0\,,\quad \Sigma=-\frac{2}{3\mu^2+8}\,,\nonumber \\
\hspace{-0.7cm}
& &
\Omega_B=\frac{3}{2(3\mu^2+8)}\,,\quad 
\Omega_r=0\,,
\quad \Omega_m=\frac{3\mu^2+6}{3\mu^2+8}\,,
\label{b3es}
\ea
with $\Omega_{\rm DE}=2/(3\mu^2+8)$ and 
$w_{\rm DE}=0$. 
This corresponds to an anisotropic scaling solution 
realizing the matter dominance for $\mu \gg 1$.
The CMB constraint 
$\Omega_{\rm DE}<0.02$
gives the bound $\mu>5.5$.

\subsubsection{Dark energy dominance}

$\bullet$ (\itl{c1}) Isotropic dark energy  dominated point
\vspace{0.1cm}
\ba
& &
x_1=\frac{\lambda}{\sqrt{6}}\,,\quad 
x_2=\sqrt{1-\frac{\lambda^2}{6}}\,,\quad 
\Sigma=0\,,\nonumber \\
& &
\Omega_B=0\,,\quad \Omega_r=0\,,
\quad \Omega_m=0\,,
\ea
with $\Omega_{\rm DE}=1$ and  $w_{\rm DE}=w_{\rm eff}=-1+\lambda^2/3$.
The condition for the cosmic acceleration ($w_{\rm eff}<-1/3$) 
corresponds to $\lambda^2<2$.

\vspace{0.2cm}

$\bullet$ (\itl{c2}) Anisotropic dark energy  dominated  point
\vspace{0.1cm}
\ba
& &
x_1=\frac{(2\lambda+\mu)\sqrt{6}}{2\lambda^2+5\lambda \mu+3\mu^2+8}\,,
\nonumber \\
& &
x_2=\frac{\sqrt{3(\lambda \mu+\mu^2+4)
(3\mu^2+4\lambda \mu+8)}}
{2\lambda^2+5\lambda \mu+3\mu^2+8}\,,\nonumber \\
& &
\Sigma=-\frac{2(\lambda^2+\lambda \mu-2)}
{2\lambda^2+5\lambda \mu+3\mu^2+8}\,,\nonumber \\
& &
\Omega_B=\frac{3(3\mu^2+4\lambda \mu+8)(\lambda^2+\lambda \mu-2)}
{(2\lambda^2+5\lambda \mu+3\mu^2+8)^2}\,,\nonumber \\
& &
\Omega_r=0\,,
\quad \Omega_m=0\,,
\label{c2es1}
\ea
with $\Omega_{\rm DE}=1$ and 
\be
w_{\rm DE}=w_{\rm eff}=
-1+\frac{2\lambda(2\lambda+\mu)}{2\lambda^2+5\lambda \mu+3\mu^2+8}\,.
\label{c2es2}
\ee
For positive values of $\lambda$ and $\mu$, $w_{\rm DE}$ 
is larger than $-1$.
Since $\Omega_B>0$, we require that 
\be
\lambda^2+\lambda \mu-2>0\,,
\label{lamcon}
\ee
for the existence of point (\itl{c2}). 
Under this condition, $\Sigma$ is negative.
The cosmic acceleration occurs under the condition
\be
4\lambda^2-2\lambda \mu-3\mu^2-8<0\,.
\label{lamcon2}
\ee
When $\lambda={\cal O}(1)$, this condition is 
well satisfied for $\mu \gg 1$. 

\subsection{Stability of fixed points}

The stability of fixed points $(x_1, x_2, \Sigma, \Omega_B, \Omega_r)$ derived 
in Sec.~\ref{fixedsec} is known by considering homogeneous 
perturbations ${\bm X}=(\delta x_1, \delta x_2, \delta \Sigma, 
\delta \Omega_B, \delta \Omega_r)$ around them. Perturbing Eqs.~(\ref{auto1})-(\ref{auto5}) up to linear order, the perturbations ${\bm X}$ obey the differential equations,  
\be
{\bm X}'={\cal M} {\bm X}\,,
\ee
where ${\cal M}$ is a $5 \times 5$ Jacobian matrix. 
The signs of eigenvalues $\nu_{1,2,3,4,5}$ of 
${\cal M}$ determine the stability of fixed points.
A fixed point is stable when all the eigenvalues are
negative (including the case of negative real parts).
If at least one of the eigenvalues is positive with 
others negative, it is called a saddle. 
If all the eigenvalues are positive, the fixed point 
is called an unstable node.

We present the eigenvalues $\nu_{1,2,3,4,5}$ of matrix 
${\cal M}$ for the fixed points obtained in Sec.~\ref{fixedsec}. 

\vspace{0.2cm}

$\bullet$ (\itl{a1})
\be	
1,\quad 2,\quad 2,\quad -1,\quad -1.
\label{a1ei}
\ee

$\bullet$ (\itl{a2})
\be
1, \quad \frac{2(\lambda+2\mu)}{\lambda}\,,\quad -1, 
\quad
-\frac{1}{2} \pm \frac{\sqrt{64-15\lambda^2}}{2\lambda}\,.
\label{a2ei}
\ee

$\bullet$ (\itl{a3})
\be
1, \quad \frac{6 \mu^2+3\lambda \mu+16}
{3\mu^2+8},\quad -1,  \quad
-\frac{1}{2} \pm 
\frac{1}{2} \sqrt{\frac{-3(7\mu^2+8)}{3\mu^2+8}}.
\label{a3ei}
\ee

$\bullet$ (\itl{b1})
\be	
1,\quad \frac{3}{2},\quad -1,\quad -\frac{3}{2},
\quad -\frac{3}{2}\,.
\label{b1ei}
\ee

$\bullet$ (\itl{b2})
\be
\frac{\lambda+3\mu}{\lambda},\quad -1,\quad -\frac{3}{2},
\quad 
-\frac{3}{4}\pm \frac{3\sqrt{24-7\lambda^2}}{4\lambda}\,.
\label{b2ei}
\ee

$\bullet$ (\itl{b3})
\be
\frac{3(3\mu^2+\lambda \mu+8)}{2(3\mu^2+8)},\quad
-1, \quad -\frac{3}{2}, \quad 
-\frac{3}{4} \pm \frac{3}{4} 
\sqrt{\frac{-(5\mu^2+8)}{3\mu^2+8}}.
\label{b3ei}
\ee

$\bullet$ (\itl{c1})
\be
\lambda^2+\lambda \mu-2,\quad 
\lambda^2-4, \quad \lambda^2-3,\quad 
\frac{\lambda^2}{2}-3,\quad 
\frac{\lambda^2}{2}-3.
\label{c1ei}
\ee

$\bullet$ (\itl{c2})
\ba
& &
\frac{3(2\lambda^2-3\lambda \mu-3\mu^2-8)}{2\lambda^2+5\lambda \mu+3\mu^2+8}\,,\quad 
-\frac{3(3\mu^2+4\lambda \mu+8)}{2\lambda^2+5\lambda \mu+3\mu^2+8}\,,\nonumber \\
& &
\frac{2(2\lambda^2-7\lambda \mu-6\mu^2-16)}{2\lambda^2+5\lambda \mu+3\mu^2+8}\,,\nonumber \\
& &
-\frac{3(3\mu^2+4\lambda \mu+8)}{2(2\lambda^2+5\lambda \mu+3\mu^2+8)} \left( 1\pm \sqrt{1-{\cal F}} \right)\,,
\label{c2ei}
\ea
where 
\be
{\cal F} \equiv \frac{4(\mu^2+\lambda \mu+4)
(\lambda^2+\lambda \mu-2)}
{3\mu^2+4\lambda \mu+8}\,.
\label{calF}
\ee

The point (\itl{a1}) is a saddle with three positive eigenvalues. 
Under the BBN bounds on $\lambda$ and $\mu$, both 
(\itl{a2}) and (\itl{a3}) are saddles with two positive eigenvalues. 
The point (\itl{b1}) is a saddle with two positive eigenvalues. 
Under the CMB bound $\lambda>12$, the point (\itl{b2}) 
is a saddle with one positive eigenvalue, while the other four eigenvalues 
are negative or have negative real parts.
The point (\itl{b3}) is also a saddle with two real 
negative eigenvalues and two complex eigenvalues with 
negative real parts. 

%%%%%%%%%%%%%%%%%%%%%%%%%%%%%%
\begin{figure}[h]
\begin{center}
\includegraphics[width=0.92\linewidth]{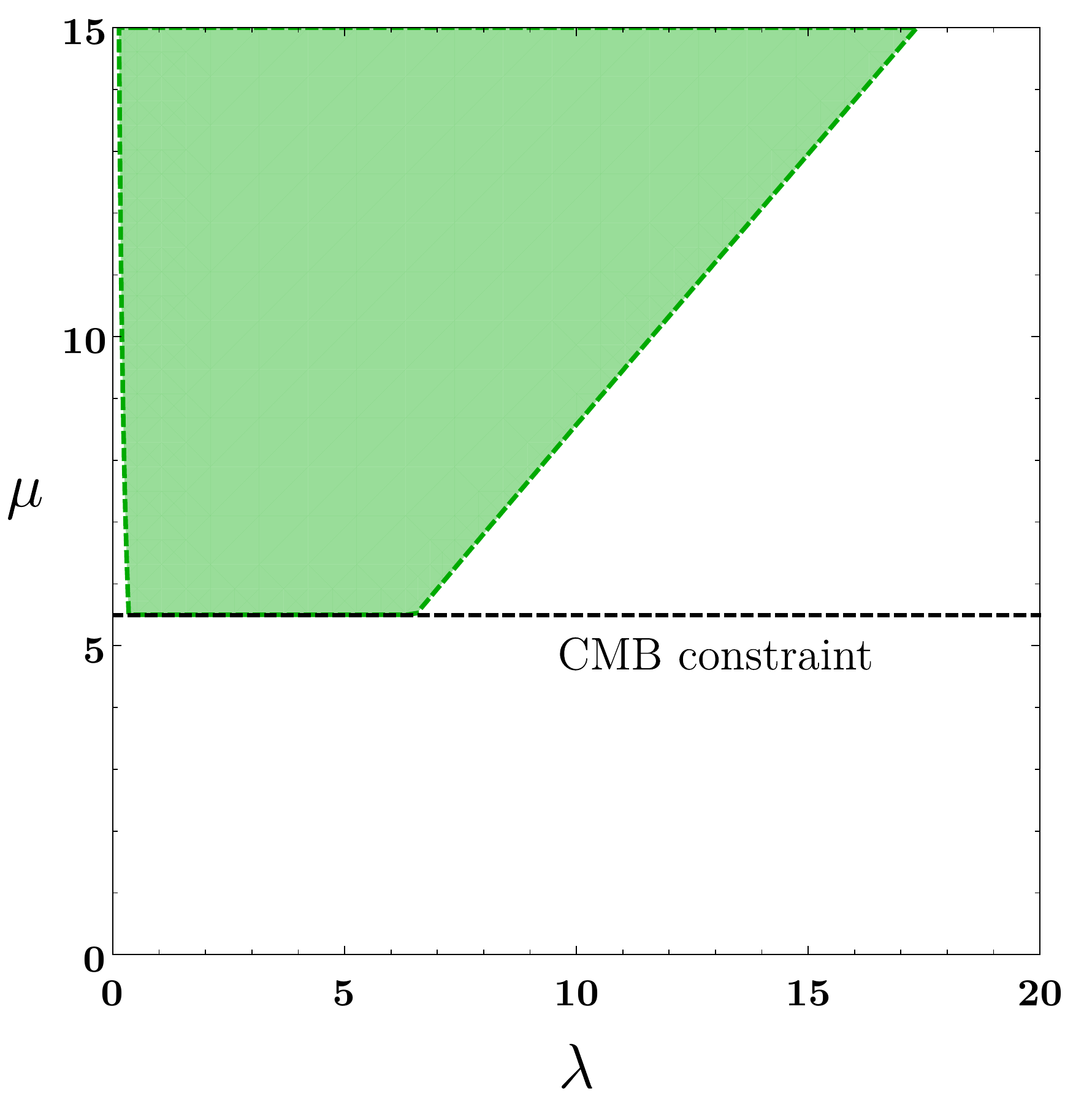}
\end{center}
\caption{\label{fig1} 
Parameter space in which the 
conditions (\ref{lamcon}) and (\ref{lamcon2}) are 
satisfied (colored region).  
If the anisotropic point (\itl{b3}) is responsible for the 
matter era, there is also the bound $\mu>5.5$ arising from 
the CMB constraint $\Omega_{\rm DE}<0.02$ 
around the redshift 50
(dashed black line). 
The variable $\mu$ is unbounded from above.}
\end{figure}
%%%%%%%%%%%%%%%%%%%%%%%%%%%%%%

The point (\itl{c1}) is responsible for the cosmic acceleration for 
$\lambda^2<2$. Under this condition, 
the point (\itl{c1}) is a saddle for 
\be
\lambda^2+\lambda \mu-2>0\,,
\label{lamsta}
\ee
whereas it is stable for $\lambda^2+\lambda \mu-2<0$. 
The condition (\ref{lamsta}) is identical to (\ref{lamcon}). 
This means that, as long as the anisotropic point (\itl{c2}) is present, 
the isotropic point (\itl{c1}) is a saddle. 
Under the condition (\ref{lamsta}), the last two eigenvalues of 
point (\itl{c2}) in Eq.~(\ref{c2ei}) are negative or complex 
with negative real parts. 
Moreover, under the condition (\ref{lamcon2}) for the cosmic 
acceleration of point (\itl{c2}), the other three eigenvalues 
in Eq.~(\ref{c2ei}) are negative. 
Provided that the two inequalities (\ref{lamcon}) and (\ref{lamcon2}) hold, 
the anisotropic dark energy  dominated point (\itl{c2}) is stable, while (\itl{c1}) is a saddle.

In Fig.~\ref{fig1}, we show the parameter space in 
the ($\lambda$, $\mu$) plane consistent with the 
conditions (\ref{lamcon}) and (\ref{lamcon2}). 
We also plot the bound $\mu>5.5$ arising from the 
CMB constraint on point (\itl{b3}). 
For the model parameters inside the colored region of 
Fig.~\ref{fig1}, the saddle anisotropic matter point (\itl{b3}) 
can be followed by the accelerated attractor (\itl{c2}) 
with the non-vanishing anisotropic shear. 

%%%%%%%%%%%%%%%%
\section{Cosmological dynamics}
\label{sec:cossec}
%%%%%%%%%%%%%%%%

We study the cosmological dynamics for the coupling 
constants $\lambda$ and $\mu$ inside the colored region 
of Fig.~\ref{fig1}. 
Prior to the radiation-dominated epoch, 
we assume the existence of 
an inflationary period (with subsequent reheating) 
driven by a scalar degree of freedom other than $\phi$. 
As long as such an additional scalar degree of freedom does not 
have specific couplings to form fields, the anisotropic shear
quickly decreases during inflation. 
Then, the natural initial condition for the anisotropic 
shear at the onset of radiation era is 
$|\Sigma|$ very close to 0.
In this case, the fixed points relevant to the early 
radiation era correspond to either (\itl{a1}) or (\itl{a2}). 
Indeed, we would like to show that, even if the 
initial condition of shear at the beginning of radiation era 
is close to the isotropic one ($\Sigma \simeq 0$), 
the solutions can approach fixed points with 
anisotropic hairs in the late Universe. 

\subsection{Sequence of fixed points}

For $\lambda>9.4$, the point (\itl{a2}) can be responsible for  
the scaling radiation era consistent with the BBN bound. 
If the coupling $f(\phi)$ is absent, it is known that 
(\itl{a2}) is followed by the isotropic matter scaling solution 
(\itl{b2}) by reflecting the fact that the latter is stable for 
$\lambda^2>3$ \cite{CLW,CST,Tsuji13}. 
This property does not hold for the theories with 
$f(\phi) \neq 0$, since the point (\itl{b2}) is a saddle. 
Instead, the point (\itl{c2}) is stable for $\lambda$ and $\mu$ 
inside the colored region of Fig.~\ref{fig1}.
Our numerical calculations show that, for the initial conditions 
close to point (\itl{a2}) during the radiation dominance 
with $\lambda>9.4$, 
the solutions directly approach point (\itl{c2}) without passing 
through the scaling matter point (\itl{b2}). 
This means the absence of a proper matter era, so the viable 
cosmological trajectory does not arise from the isotropic 
radiation scaling solution (\itl{a2}).

The initial conditions in the deep radiation era realizing the viable 
late-time cosmology are those close to the isotropic radiation point (\itl{a1}). 
Then, the isotropic scaling solutions (\itl{a2}) and (\itl{b2}) are 
irrelevant to the cosmological dynamics in the following discussion. 
We recall that the anisotropic radiation point (\itl{a3}) has one less 
positive eigenvalues of matrix ${\cal M}$ than those of  (\itl{a1}). 
This suggests that the solutions may temporally approach point (\itl{a3}) 
during the late radiation era. This is indeed the case for 
numerical analysis presented later.

The point (\itl{b1}) has three negative eigenvalues of 
matrix ${\cal M}$, while point (\itl{b3}) has two negative 
eigenvalues and two complex eigenvalues with negative real parts.  
Then, after the radiation dominance, the solutions should temporally 
approach the anisotropic matter point (\itl{b3}) 
rather than the isotropic matter point (\itl{b1}).
As we mentioned in Sec.~\ref{fixedsec}, the point (\itl{b3}) is 
consistent with the CMB bound $\Omega_{\rm DE}<0.02$ 
around the redshift $50$ for
\be
\mu>5.5\,,
\label{mura}
\ee
whose condition is imposed in the following.
Since point (\itl{b3}) is a saddle, the solutions eventually 
exit from the matter era driven by  (\itl{b3}) to reach
the stable anisotropic point (\itl{c2}) with cosmic acceleration.

%%%%%%%%%%%%%%%%%%%%%%%%%%%%%%
\begin{figure}[h]
\begin{center}
\includegraphics[width=0.92\linewidth]{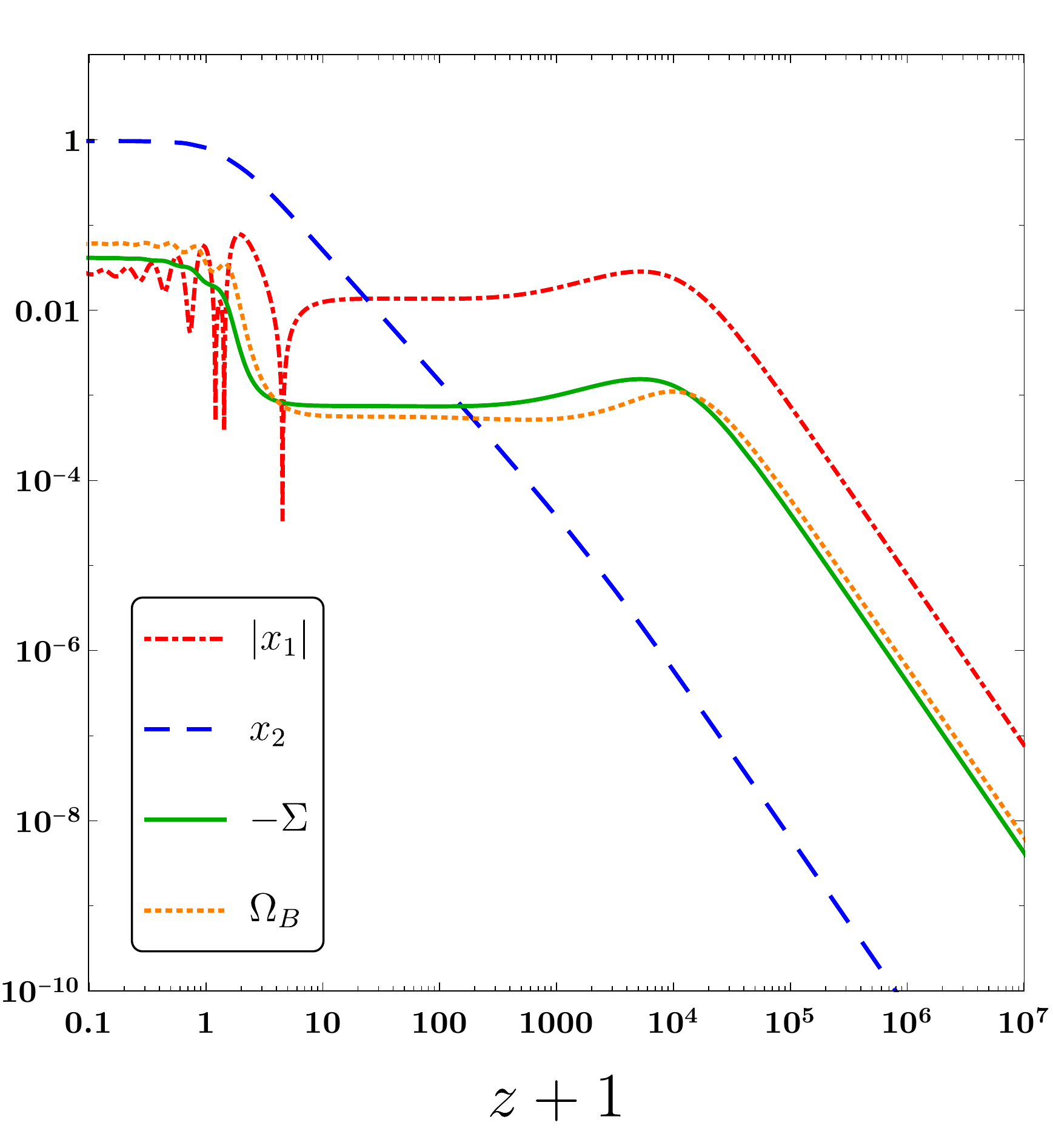}
\includegraphics[width=0.92\linewidth]{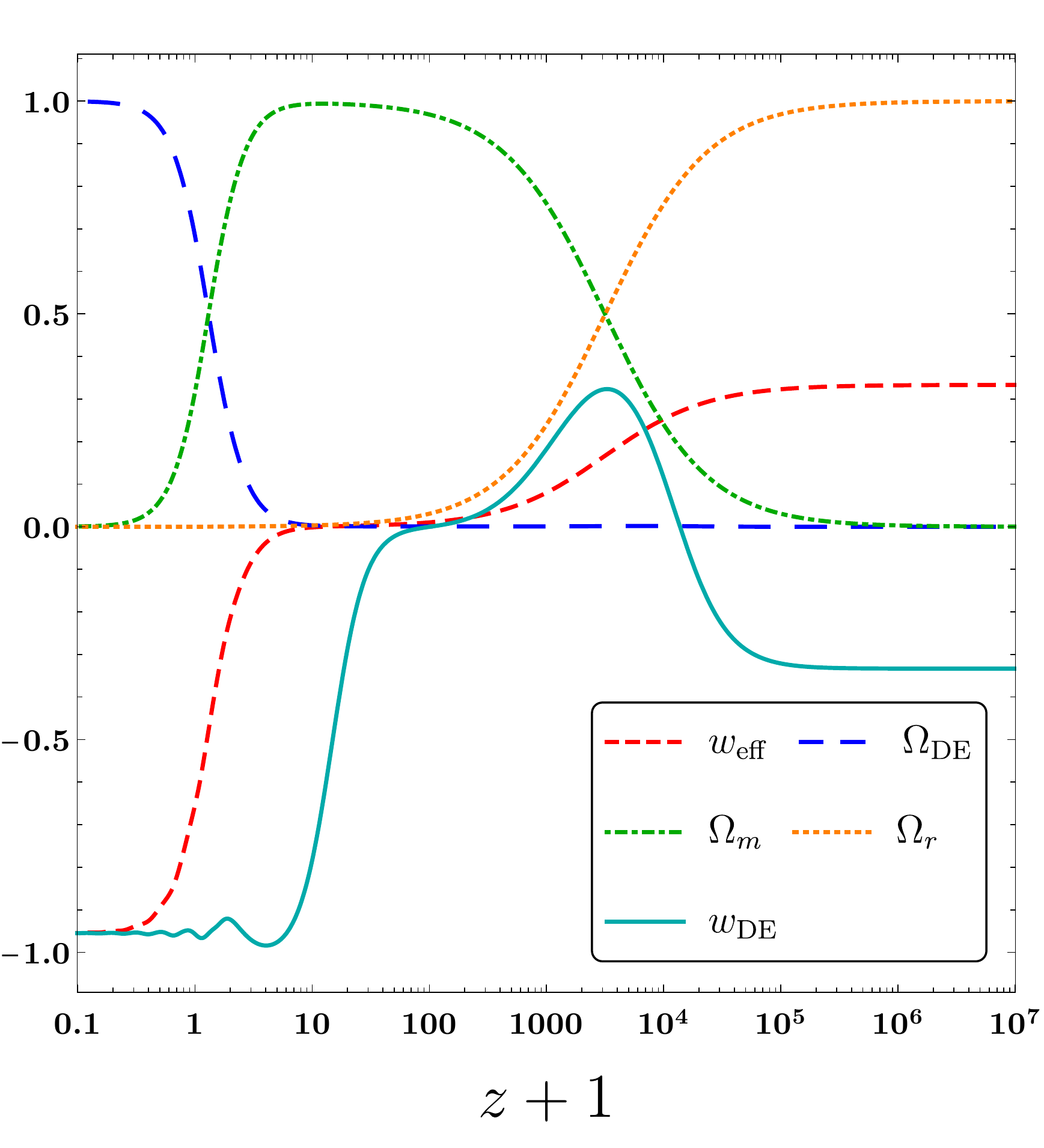}
\end{center}
\caption{\label{fig2}
Evolution of $|x_1|$, $x_2$, $-\Sigma$, $\Omega_B$ (top) 
and $\Omega_{\rm DE}$, $\Omega_r$, $\Omega_m$, 
$w_{\rm DE}$, $w_{\rm eff}$ (bottom) versus $z+1~(=1/a)$ 
for $\lambda=2$ and $\mu=30$ with 
the initial conditions $x_1=10^{-13}$, $x_2=10^{-14}$, $\Sigma=0$, $\Omega_B=10^{-10}$, and $\Omega_r=0.99996$ at the redshift $z=7.9 \times 10^7$. 
The present epoch (redshift $z=0$) 
is identified by $\Omega_{\rm DE}=0.68$.
}
\end{figure}
%%%%%%%%%%%%%%%%%%%%%%%%%%%%%%

In Fig.~\ref{fig2}, we show the numerical solutions to $|x_1|$, $x_2$, $-\Sigma$, $\Omega_B$ 
as well as  $\Omega_{\rm DE}$, $\Omega_r$, $\Omega_m$, 
$w_{\rm DE}$, $w_{\rm eff}$ derived 
by numerically integrating Eqs.~(\ref{auto1})-(\ref{auto5})
for $\lambda=2$ and $\mu=30$. 
The initial values of $x_1$, $x_2$, $\Omega_B$
are very much smaller than 1, so the solutions 
start from the regime close to the isotropic radiation 
point (\itl{a1}). The initial condition of $\Sigma$ 
is chosen to be 0, but the cosmological dynamics 
hardly changes for $|\Sigma|$ initially much smaller than 1. 
The existence of non-zero $\Omega_B$ is crucial to generate 
the non-vanishing anisotropic shear at late times. 
As we will see below, the tiny initial value of  
$\Omega_B$ like the order $10^{-10}$ is
sufficient for achieving this purpose. 
In Fig.~\ref{fig2}, the condition $\Omega_B \gg \{ x_1^2, x_2^2, \Sigma^2 \}$ is 
satisfied in the early radiation era, so the dark energy equation of state (\ref{wde}) 
is close to $w_{\rm DE}=-1/3$ 
during this epoch (see the bottom panel of Fig.~\ref{fig2}).

In Fig.~\ref{fig2}, the 
radiation-dominated epoch ($\Omega_r \simeq 1$ and 
$w_{\rm eff} \simeq 1/3$) is followed by the 
matter-dominated era ($\Omega_m \simeq 1$ and 
$w_{\rm eff} \simeq 0$) around the redshift 
$z \equiv 1/a-1 \simeq 3200$. 
The variables $|x_1|$, $x_2$, $-\Sigma$, $\Omega_B$ increase 
during the deep radiation era. 
After the transient period in which $w_{\rm DE}$ increases 
from $-1/3$ to the value close to $1/3$, the solutions enter 
the stage in which $|x_1|$, $-\Sigma$, $\Omega_B$ 
are nearly constant. The increase of $w_{\rm DE}$ continues 
by the radiation-matter equality. 
This behavior of $w_{\rm DE}$ can be interpreted as the temporal approach 
to the anisotropic radiation point (\itl{a3}) characterized by 
$w_{\rm DE}=1/3$. Indeed, the numerical values of  $|x_1|$, $\Sigma$, 
$\Omega_B$ around $z=3200$ are in fairly good agreement with 
their analytic values computed from Eq.~(\ref{a3}). 
In other words, the anisotropic shear of order $\Sigma=-4/(3\mu^2+8)$ 
is already generated around the end of radiation era. 

We note that the moment at which the transition from 
(\itl{a1}) to (\itl{a3}) takes place depends on the initial 
values of $x_1$, $x_2$, $\Omega_B$. 
We numerically find that there are cases in which the transition 
occurs much earlier compared to Fig.~\ref{fig2}. 
In such cases, the solutions stay around 
the point (\itl{a3}) characterized by 
$w_{\rm DE}=1/3$ for a longer period 
during the radiation era.

The regime in which the variables $|x_1|$, $-\Sigma$, $\Omega_B$ stay 
nearly constant after the radiation-matter equality corresponds to the 
anisotropic scaling matter fixed point (\itl{b3}).
{}From Eq.~(\ref{b3es}), we have 
$x_1=-1.36 \times 10^{-2}$, $\Sigma=-7.39 \times 10^{-4}$,
and $\Omega_B=5.54 \times 10^{-4}$ on point (\itl{b3}), 
which exhibit good agreement with their numerical values around $z=60$. 
The dark energy density parameter on point (\itl{b3}) is given 
by $\Omega_{\rm DE}=7.39 \times 10^{-4}$, which 
is consistent with the CMB bound $\Omega_{\rm DE}(z=50)<0.02$.
In the bottom panel of Fig.~\ref{fig2}, we can confirm that 
the solutions temporally reach the region around 
$w_{\rm DE}=0$ during the matter era (which corresponds 
to the value of $w_{\rm DE}$ on point (\itl{b3})).
 
In the top panel of Fig.~\ref{fig2}, we observe that 
$x_2$ exceeds $|x_1|$ around $z=22$. 
This signals the departure from the anisotropic matter 
fixed point (\itl{b3}). Indeed, $w_{\rm DE}$ starts to deviate 
from 0 for $z \lesssim 30$. 
The anisotropic dark energy  dominated point (\itl{c2}) is stable 
for $\lambda=2$ and $\mu=30$, while the isotropic  
point (\itl{c1}) is not. As we see in Fig.~\ref{fig2},  
the solutions finally approach the fixed point (\itl{c2}) with 
the non-vanishing anisotropic shear after 
the matter-dominated epoch.
{}From Eqs.~(\ref{c2es1}) and (\ref{c2es2}), 
we have $x_1=2.76 \times 10^{-2}$, $x_2=0.968$, 
$\Sigma=-4.11 \times 10^{-2}$, $\Omega_B=6.03 \times 10^{-2}$, 
and $w_{\rm DE}=w_{\rm eff}=-0.955$ on point (\itl{c2}), 
which are in good agreement with their numerical values 
in the asymptotic future ($z \to -1$). 
The potential energy $V(\phi)$, which is associated with 
the variable $x_2$, is the main source for $\Omega_{\rm DE}$ 
at late times, but the 2-form energy density characterized by 
$\Omega_B$ also provides the non-negligible 
contribution to $\Omega_{\rm DE}$. 

Since $x_1<0$ and $x_1>0$ on points (\itl{b3}) and (\itl{c2}), respectively, 
$x_1$ changes its sign during the transition from the end of matter era 
to the dark energy  dominated epoch (around $z=3.5$ in Fig.~\ref{fig2}).
The quantity $\Sigma$, which is negative, survives 
during the cosmological sequence of  
$(\itl{a3}) \to (\itl{b3}) \to (\itl{c2})$. 
Since the 2-form energy density $\rho_B$ is the source for the anisotropic 
shear, $\Omega_B$ evolves in the similar way to $-\Sigma$. 
We note that the condition $\Sigma^2 \ll \Omega_B$ is always 
satisfied in the numerical integration of Fig.~\ref{fig2}, 
so we can ignore the terms $\Sigma^2$ in 
Eqs.~(\ref{Omede}) and (\ref{wde}).

In the bottom panel of Fig.~\ref{fig2}, we find that $w_{\rm DE}$ temporally 
reaches the minimum value $-0.984$ around $z=3$ and then it finally 
approaches the asymptotic value $-0.955$ with oscillations. 
The quantity ${\cal F}$ in Eq.~(\ref{calF}) 
is larger than 1 for $\lambda=2$ and $\mu=30$, 
so two of the eigenvalues of matrix ${\cal M}$ 
in Eq.~(\ref{c2ei}) are complex with negative real parts. 
In this case the point (\itl{c2}) is a stable spiral, so the oscillation 
of $w_{\rm DE}$ occurs before reaching the attractor. 
More generally, point (\itl{c2}) is the stable spiral 
for ${\cal F}>1$. This condition translates to
\be
4\lambda \mu^3+(8\lambda^2-11) \mu^2
+4\lambda (\lambda^2+1)\mu
+8(2\lambda^2-5)>0\,.
\label{lammu2}
\ee
When $\lambda=1$, for example, this inequality gives $\mu>1.68$. 

For the couplings satisfying $\mu \gg \lambda \geq {\cal O}(1)$, 
the dark energy equation of state 
(\ref{c2es2}) on point (\itl{c2}) is approximately given by 
\be
w_{\rm DE} \simeq -1-\Sigma\,,
\label{wdeap}
\ee
where 
\be
\Sigma \simeq -\frac{2\lambda}{3\mu}\,.
\label{Siges}
\ee
In the limit $\lambda/\mu \to 0$, we have $w_{\rm DE} \to -1$ 
with $\Sigma \to 0$. This is consistent with the no-hair theorem 
on the de Sitter background \cite{Wald,Starobinsky1982mr}.
The coupling $f(\phi)$ in the range $0<\lambda/\mu \ll 1$
allows the possibility for realizing the late-time cosmic acceleration 
with the surviving anisotropic hair. If the coupling $f(\phi)$ is absent, 
the accelerated expansion occurs only for $\lambda<\sqrt{2}$.
The numerical solution in Fig.~\ref{fig2} shows that $w_{\rm DE}$ 
close to $-1$ can be realized at low redshifts even for $\lambda>\sqrt{2}$.
We also note that, for $\mu \gg \lambda \geq {\cal O}(1)$, the condition
(\ref{lammu2}) is always satisfied, so the solutions finally approach the 
stable spiral point (\itl{c2}) with the oscillation of $w_{\rm DE}$.

%%%%%%%%%%%%%%%%%%%%%%%%%%%%%%
\begin{figure}[h]
\begin{center}
\includegraphics[width=\linewidth]{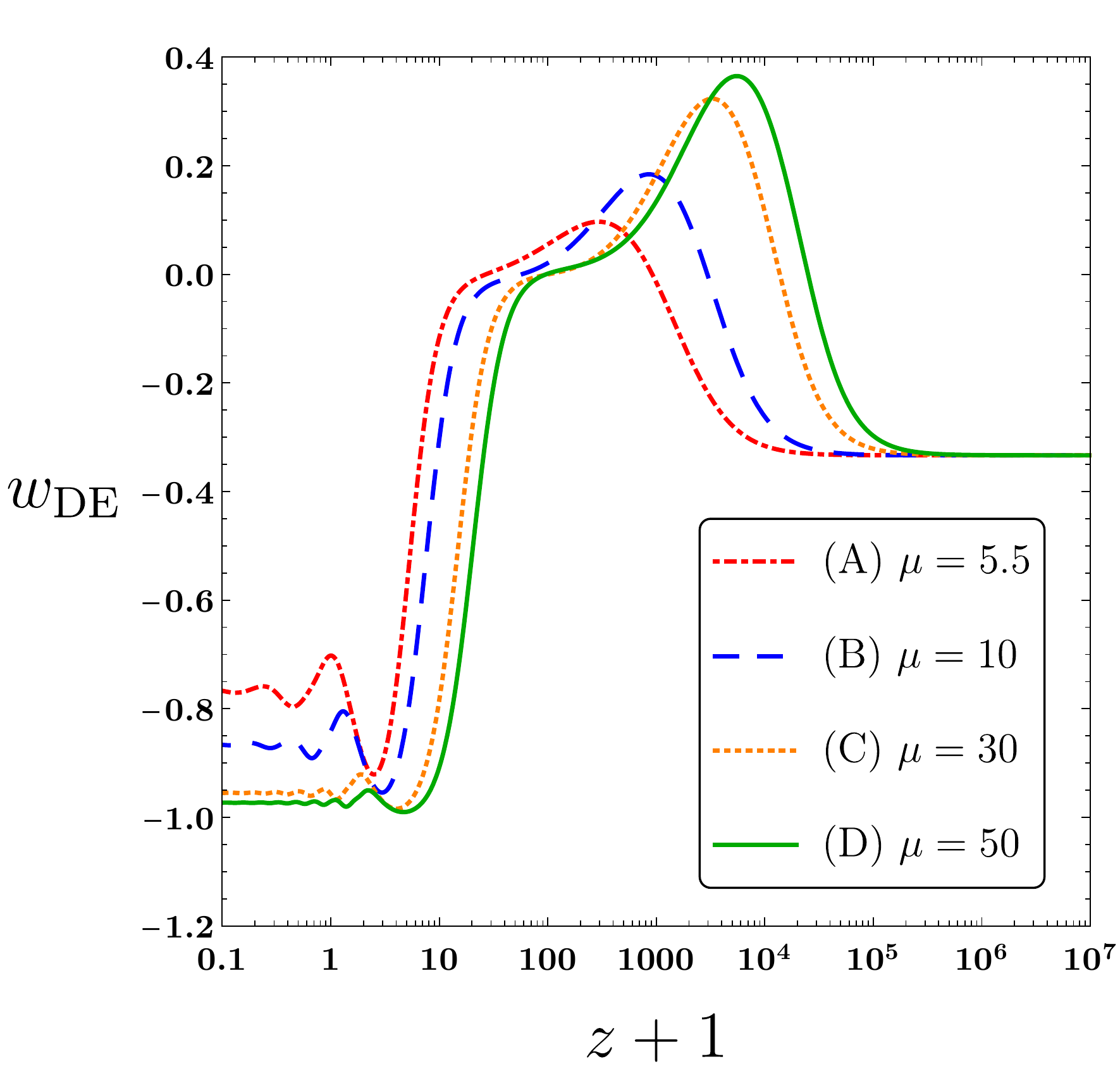}
\end{center}
\caption{\label{fig3}
Evolution of $w_{\rm DE}$ versus $z+1$ for 
$\lambda=2$ with the same initial values of 
$x_1$, $x_2$, $\Sigma$, and $\Omega_B$, 
as those used in Fig.~\ref{fig2}.
Each line corresponds to 
(A) $\mu=5.5$, (B) $\mu=10$, (C) $\mu=30$, and 
(D) $\mu=50$. 
The initial conditions of radiation density parameter 
are chosen to be 
(A) $\Omega_r=0.99994$ at  $z=5.5 \times 10^{7}$, 
(B) $\Omega_r=0.999951$ at  $z=6.5 \times 10^{7}$, 
(C) $\Omega_r=0.99996$ at  $z=7.9 \times 10^{7}$, and 
(D) $\Omega_r=0.999961$ at  $z=8.3 \times 10^{7}$, 
respectively, 
to realize the value $\Omega_r(z=0) \simeq 10^{-4}$.
}
\end{figure}
%%%%%%%%%%%%%%%%%%%%%%%%%%%%%%

In Fig.~\ref{fig3}, we plot the evolution of $w_{\rm DE}$ for 
four different values of $\mu$ by fixing $\lambda$ to be 2. 
As the analytic estimation (\ref{wdeap}) shows, for increasing $\mu$, 
the future asymptotic values of $w_{\rm DE}$ decrease
toward $-1$. In case (A), i.e., $\mu=5.5$, the CMB bound 
$\Omega_{\rm DE}(z=50)<0.02$ is marginally satisfied, 
with $w_{\rm DE}=-0.70$ today. This case should be in tension 
with observational bounds on $w_{\rm DE}$. 
For larger $\mu$, however, the values of 
$w_{\rm DE}$ at low redshifts get smaller. 
In cases (B), (C), (D) of Fig.~\ref{fig3}, today's values of 
$w_{\rm DE}$ are $-0.84$, $-0.96$, and $-0.97$, respectively. 
Moreover, for larger $\mu$, $w_{\rm DE}$ reaches the minima 
closer to $-1$ at earlier cosmological epochs.

\subsection{Observational signatures}

{}From the magnitude-redshift data of SN Ia measurements, 
the analysis of Ref.~\cite{Campanelli:2010zx} based on an 
anisotropic fluid showed that today's value of $\Sigma$ is 
constrained to be $|\Sigma (t_0)| \leq {\cal O}(0.01)$.
For the model parameters plotted in Fig.~\ref{fig2}, 
$-\Sigma (t_0)$ is of order $0.01$. 
{}From Eq.~(\ref{Siges}), today's value of $|\Sigma (t_0)|$ 
decreases further for the smaller ratio $\lambda/\mu$, 
in which case the model should be well within the SN Ia bound.

%%%%%%%%%%%%%%%%%%%%%%%%%%%%%%
\begin{figure}[h]
\begin{center}
\includegraphics[width=0.9\linewidth]{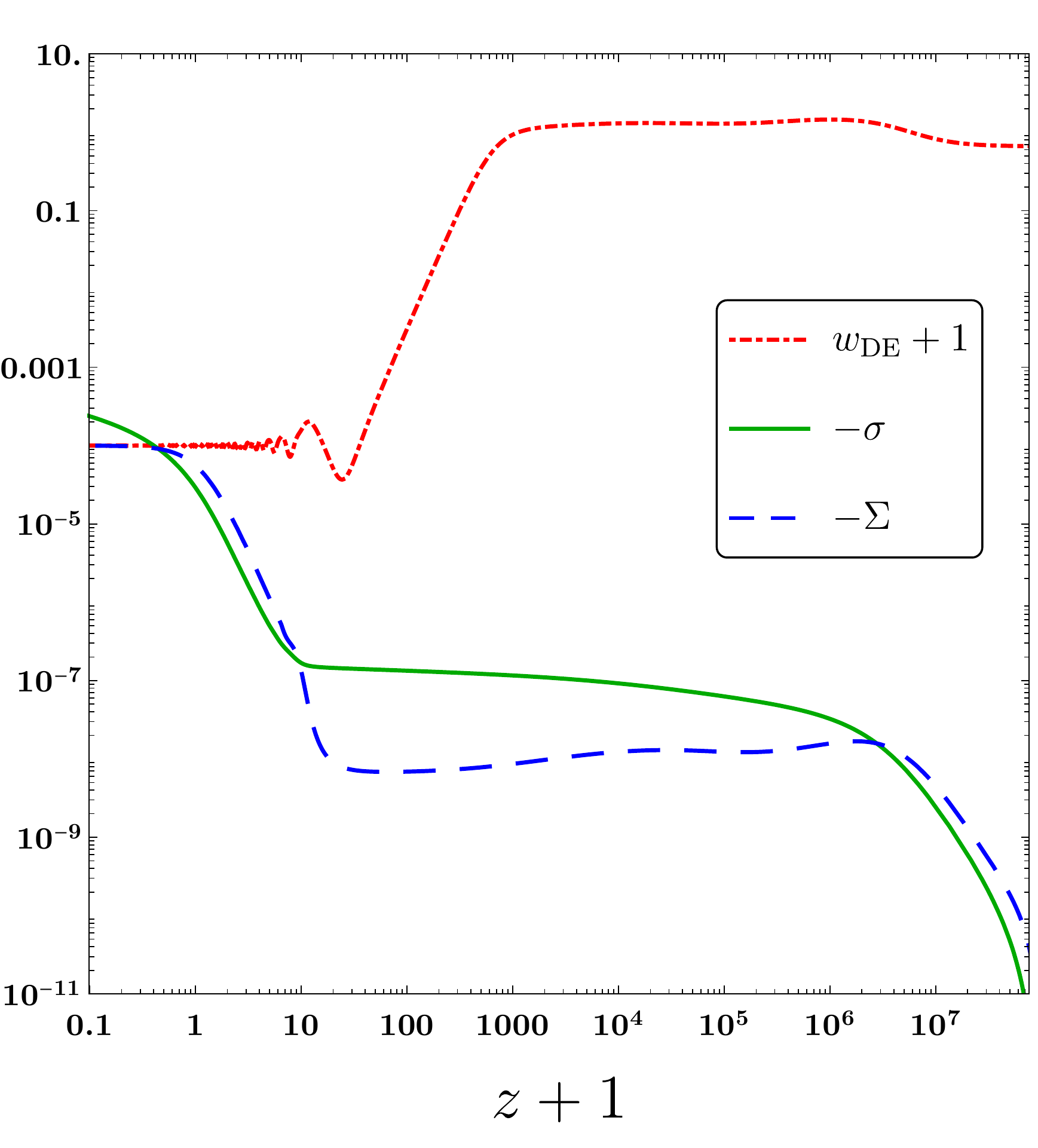}
\end{center}
\caption{\label{fig4}
Evolution of $-\sigma$, $-\Sigma$, and $w_{\rm DE}+1$ 
versus $z+1$ for $\lambda=1.5$ and $\mu=10^4$. 
The initial conditions are chosen to be 
$x_1=10^{-13}$, $x_2=10^{-14}$, 
$\Sigma=0$, $\Omega_B=10^{-10}$, 
$\Omega_r=0.999965$, and $\sigma=0$ 
at the redshift $z=9.1 \times 10^7$. }
\end{figure}
%%%%%%%%%%%%%%%%%%%%%%%%%%%%%%

The time variation of $\sigma$ after decoupling also 
leads to the modification to CMB temperature 
anisotropies \cite{Koivisto:2007bp,Koivisto:2008ig}.
The spatial metric tensor $g_{ij}$ for the line element 
(\ref{anisotropic-metric}) is expressed in the form 
$g_{ij}=a^2(t) \gamma_{ij}$, where $\gamma_{ij}$ 
is the anisotropic contribution containing $\sigma(t)$. 
Defining $\sigma_{ij} \equiv \dot{\gamma}_{ij}/2$, 
the CMB temperature anisotropy due to the anisotropic 
shear is quantified as \cite{Appleby:2009za}
\be
\frac{\delta T}{T} (\hat{n})=-\int_{t_{\rm de}}^{t_0} 
\sigma_{ij} \hat{n}^i \hat{n}^j \rm{d}t\,,
\ee
where $t_{\rm de}$ and $t_0$ correspond to the cosmic time 
at decoupling ($z \simeq 1090$) and today ($z=0$), 
respectively, and $\hat{n}$ is the 
line-of-sight unit vector. The scalar product 
$\sigma_{ij} \hat{n}^i \hat{n}^j$ 
is at most of order $\dot{\sigma}$ and hence 
\be
\left| \frac{\delta T}{T} (\hat{n})\right| 
\lesssim \left| \int_{t_{\rm de}}^{t_0} 
\dot{\sigma}\,\rm{d}t \right|
=\left| \sigma(t_0)-\sigma(t_{\rm de}) 
\right| \,.
\label{delTes}
\ee
Provided that the right hand side of Eq.~(\ref{delTes}) 
is much smaller than 1, the anisotropic shear mostly 
affects the CMB quadrupole. 
According to the analysis of Ref.~\cite{Appleby:2009za}, 
the conservative criterion for the consistency with the CMB quadrupole data
should be around $\left| \sigma(t_0)-\sigma(t_{\rm de}) \right|<10^{-4}$.

Since $\sigma'=\Sigma$, the quantity $\left| \sigma(t_0)-\sigma(t_{\rm de}) \right|$ 
is related to the asymptotic value $\Sigma \simeq -2\lambda/(3\mu)$ on point (\itl{c2}) 
and the other value $\Sigma=-2/(3\mu^2+8)$ on point (\itl{b3}). 
In Fig.~\ref{fig4}, we plot the evolution of $-\sigma$ and $-\Sigma$ 
for $\lambda=1.5$ and $\mu=10^4$ by choosing their initial conditions to be 0. 
As estimated analytically, $-\Sigma$ temporally reaches 
the nearly constant value $6.7 \times 10^{-9}$ in the matter era 
and finally approaches the asymptotic value $1.0 \times 10^{-4}$. 
In this case, the numerical values of $-\sigma$ today and at decoupling 
are given, respectively, by $-\sigma(t_0)=2.8 \times 10^{-5}$ and 
$-\sigma(t_{\rm de})=1.2 \times 10^{-7}$, so that 
$\left| \sigma(t_0)-\sigma(t_{\rm de}) \right|=2.8 \times 10^{-5}<10^{-4}$.

The above discussion shows that the models in which both $\mu/\lambda$ 
and $\mu$ are much larger than the order 1 can be consistent with the CMB quadrupole bound. 
For such model parameters, the deviation of $w_{\rm DE}$ from $-1$ 
is small on the attractor point (\itl{c2}), see Eq.~(\ref{wdeap}) with Eq.~(\ref{Siges}).
In Fig.~\ref{fig4}, we observe that $w_{\rm DE}$ approaches the value 
around $-1$ at high redshifts. 
The evolution of $w_{\rm DE}$ close to $-1$ at low redshifts is 
realized by the coupling between 2-form and scalar fields 
even for the scalar exponential potential with $\lambda>\sqrt{2}$.
Moreover, our anisotropic dark energy model with
$\left| \sigma(t_0)-\sigma(t_{\rm de}) \right|={\cal O}(10^{-5})$ can leave 
interesting signatures in the CMB quadrupole anisotropy, which may be used 
to alleviate the observed quadrupole anomaly problem \cite{Aghanim:2018eyx}.

%%%%%%%%%%%%%%%%
\section{Conclusions}
\label{sec:consec}
%%%%%%%%%%%%%%%%

We proposed a novel anisotropic dark energy model in which  
a quintessence scalar $\phi$ is coupled to a 2-form field strength $H_{\alpha \beta \gamma}$ 
with the interacting Lagrangian $-f(\phi)H_{\alpha \beta \gamma} H^{\alpha \beta \gamma}/12$.
For the exponential scalar potential $V(\phi)=V_0 e^{-\lambda \phi/M_{\rm pl}}$ 
with the coupling $f(\phi)=f_0 e^{-\mu \phi/M_{\rm pl}}$, we showed that the late-time 
cosmic acceleration with the dark energy equation of state $w_{\rm DE}$ close to 
$-1$ can be realized even for $\lambda \geq {\cal O}(1)$. 
This property comes from the fact that there exists the anisotropic 
accelerated attractor fixed point (\itl{c2}) supported by the 2-form 
density parameter $\Omega_B$ and the shear $\Sigma$.

Even for initial conditions close to the isotropic radiation point (\itl{a1}), 
we showed that the solutions temporally reach the saddle anisotropic point (\itl{a3}) 
by the end of the radiation era and then they are followed by 
the saddle anisotropic matter scaling solution (\itl{b3}) with constant 
$\Sigma$ and $\Omega_B$. {}From the CMB bound 
$\Omega_{\rm DE} (z=50)<0.02$ on the scaling matter fixed point 
(\itl{b3}), the coupling constant $\mu$ is constrained to be $\mu>5.5$.
Provided that the two conditions (\ref{lamcon}) and (\ref{lamcon2}) are satisfied, 
the fixed point (\itl{c2}) corresponds to the accelerated attractor 
with non-vanishing anisotropic hair.

In summary, the typical cosmological evolution is given by the trajectory,
\be
(\itl{a1}) \to  (\itl{a3}) \to (\itl{b3}) \to (\itl{c2})\,.
\ee
For the couplings in the range $\mu \gg \lambda \geq {\cal O}(1)$, 
we numerically confirmed the above cosmological sequence, see e.g., 
Fig.~\ref{fig2}. The analytic derivation of point 
(\itl{c2}) showed that, for the larger ratio $\mu/\lambda$, the future 
asymptotic values of $w_{\rm DE}$ 
tend to be smaller, which is the case for the
numerical integration in Fig.~\ref{fig3}. 
Before reaching the attractor, $w_{\rm DE}$ exhibits oscillations 
in the range $w_{\rm DE}>-1$. This property can be used to 
distinguish our model from quintessence and the $\Lambda$CDM model. 

The existence of non-vanishing anisotropic shear after the radiation-dominated 
epoch leaves imprints on observables associated with 
CMB and SN Ia measurements. 
In particular, the time variation of spatial shear $\sigma$ after decoupling 
to today affects the CMB quadrupole temperature anisotropy. 
When both $\mu/\lambda$ and $\mu$ are much larger than unity, 
we showed that the change of spatial shear from decoupling 
to today can be compatible with the CMB quadrupole data. 
In particular, if $\left| \sigma(t_0)-\sigma(t_{\rm de}) \right|$ is 
of order $10^{-5}$, there may be an interesting possibility for
addressing the problem of CMB quadrupole anomaly. 
We leave detailed observational constraints on the parameters 
$\mu$ and $\lambda$ for a future work.

%%%%%%%%%%%%%%%%%%%%%%%%%%%%%%%%%%%%%%%%%%%%
\section*{Acknowledgments}
This work was partly supported by COLCIENCIAS grant 110671250405 RC FP44842-103-2016 
and by COLCIENCIAS -- DAAD grant 110278258747 RC-774-2017. JPBA acknowledge support from 
Universidad Antonio Nari\~no grant  2017239 and thanks Tokyo University of Science for kind hospitality at early stages of this project. 
RK is supported by the Grant-in-Aid for Young Scientists B of the JSPS No.\,17K14297.  ST is supported by the Grant-in-Aid for Scientific Research Fund of the 
JSPS No.~16K05359 and MEXT KAKENHI Grant-in-Aid for Scientific Research on Innovative Areas ``Cosmic Acceleration'' (No.\,15H05890). 
%%%%%%%%%%%%%%%%%%%%%%%%%%%%%

\appendix

%%%%%%%%%%%%%%%%%%%%%%%%%%%
\section{Other fixed points}
\label{a4b4}
%%%%%%%%%%%%%%%%%%%%%%%%%%%

In this Appendix, we present two additional fixed points (\itl{a4}) and 
(\itl{b4}), which are irrelevant to the radiation, matter, dark energy  
dominated epochs.

\vspace{0.2cm}

$\bullet$ (\itl{a4}) Anisotropic scaling solution with 
$w_{\rm eff}=1/3$

\ba
& &
x_1=\frac{2\sqrt{6}}{3\lambda}\,,\quad 
x_2=\frac{\sqrt{3(6\mu^2+3\lambda\mu+16)}}
{6\lambda}\,,\notag\\ 
&&
\Sigma=-\frac{\lambda+2\mu}{2\lambda}\,,\quad 
\Omega_B=\frac{\lambda+2\mu}{4\lambda}\,,\notag\\
&&
\Omega_r=\frac{2\lambda^2-7\lambda\mu-6\mu^2-16}{4\lambda^2}\,,
\quad \Omega_m=0\,,
\ea
with $\Omega_{\rm DE}=(2\lambda^2+7\lambda\mu+6\mu^2+16)/(4\lambda^2)$ and $w_{\rm DE}=1/3$.
In the limit $\mu \to -\lambda/2$, this point reduces to 
the isotropic radiation scaling solution (\itl{a2}).
Since $w_{\rm eff}=1/3$ on point (\itl{a4}), 
it can be used only for the radiation era. 
For $\lambda$ and $\mu$ in the range (\ref{lammu}), 
however, $\Omega_r$ can not be close to 1.

\vspace{0.2cm}

$\bullet$ (\itl{b4}) Anisotropic scaling solution with 
$w_{\rm eff}=0$

\ba
& &
x_1=\frac{\sqrt{6}}{2\lambda}\,,\quad 
x_2=\frac{\sqrt{3(3\mu^2+\lambda\mu+8)}}
{4\lambda}\,,\notag\\ 
&&
\Sigma=-\frac{\lambda+3\mu}{4\lambda}\,,\quad 
\Omega_B=\frac{3(\lambda+3\mu)}{16\lambda}\,,\notag\\
&&
\Omega_r=0\,,
\quad \Omega_m=\frac{3(2\lambda^2-3\lambda\mu-3\mu^2-8)}{8\lambda^2}\,,
\ea
with $\Omega_{\rm DE}=(2\lambda^2+9\lambda\mu+9\mu^2+24)/(8\lambda^2)$ and $w_{\rm DE}=0$. 
In the limit $\mu \to -\lambda/3$, this recovers 
the isotropic matter scaling solution (\itl{b2}).
Since $w_{\rm eff}=0$ on point (\itl{b4}), 
it is relevant only to the matter era. 
For $\lambda$ and $\mu$ in the range (\ref{lammu}), however, 
$\Omega_m$ is away from 1.

%%%%%%%%%%%%%%%%%


\begin{thebibliography}{99}
%%%%%%%%%%%%%%%%%

\bibitem{SN1}
A.~G.~Riess \textit{et al.},
%``Observational evidence from supernovae
%for an accelerating universe and a cosmological constant,''
Astron.\ J.\  {\bf 116}, 1009 (1998) 
[astro-ph/9805201].

\bibitem{SN2}
S.~Perlmutter \textit{et al.},
%``Measurements of Omega and Lambda from
%42 high redshift supernovae,''
Astrophys.\ J.\  {\bf 517}, 565 (1999) 
[astro-ph/9812133].

\bibitem{Weinberg} 
S.~Weinberg,
%``The Cosmological Constant Problem,''
Rev.\ Mod.\ Phys.\  {\bf 61}, 1 (1989).
%doi:10.1103/RevModPhys.61.1

\bibitem{quin1}
Y.~Fujii, Phys.\ Rev.\ D {\bf 26}, 2580 (1982).

\bibitem{quin2}
L.~H.~Ford,
%``Cosmological Constant Damping 
%By Unstable Scalar Fields,''
Phys.\ Rev.\ D {\bf 35}, 2339 (1987).

\bibitem{quin3} 
B.~Ratra and P.~J.~E.~Peebles,
%``Cosmological Consequences of a Rolling 
%Homogeneous Scalar Field,''
Phys.\ Rev.\ D {\bf 37}, 3406 (1988).
%doi:10.1103/PhysRevD.37.3406

\bibitem{quin4}
C.~Wetterich,
%``Cosmology and the Fate of Dilatation Symmetry,''
Nucl.\ Phys.\ B {\bf 302}, 668 (1988)
%doi:10.1016/0550-3213(88)90193-9
[arXiv:1711.03844 [hep-th]].

\bibitem{quin5}
T.~Chiba, N.~Sugiyama and T.~Nakamura,
%``Cosmology with x matter,''
Mon.\ Not.\ Roy.\ Astron.\ Soc.\  {\bf 289}, L5 (1997)
[astro-ph/9704199].

\bibitem{quin6}
P.~G.~Ferreira and M.~Joyce,
%``Structure formation with a selftuning scalar field,''
Phys.\ Rev.\ Lett.\  {\bf 79}, 4740 (1997)
[astro-ph/9707286].

\bibitem{quin7}
R.~R.~Caldwell, R.~Dave and P.~J.~Steinhardt,
%``Cosmological imprint of an energy component 
%with general equation of state,''
Phys.\ Rev.\ Lett.\  {\bf 80}, 1582 (1998)
[astro-ph/9708069].

\bibitem{CLW} 
E.~J.~Copeland, A.~R.~Liddle and D.~Wands,
%``Exponential potentials and cosmological 
%scaling solutions,''
Phys.\ Rev.\ D {\bf 57}, 4686 (1998)
%doi:10.1103/PhysRevD.57.4686
[gr-qc/9711068].

\bibitem{CST} 
E.~J.~Copeland, M.~Sami and S.~Tsujikawa,
%``Dynamics of dark energy,''
Int.\ J.\ Mod.\ Phys.\ D {\bf 15}, 1753 (2006)
%doi:10.1142/S021827180600942X
[hep-th/0603057].

\bibitem{Tsuji13} 
S.~Tsujikawa,
%``Quintessence: A Review,''
Class.\ Quant.\ Grav.\  {\bf 30}, 214003 (2013)
%doi:10.1088/0264-9381/30/21/214003
[arXiv:1304.1961 [gr-qc]].

\bibitem{Garriga:2000cv} 
J.~Garriga and A.~Vilenkin,
%``Solutions to the cosmological constant problems,''
Phys.\ Rev.\ D {\bf 64}, 023517 (2001)
%doi:10.1103/PhysRevD.64.023517
[hep-th/0011262].

\bibitem{Emparan:2003gg} 
R.~Emparan and J.~Garriga,
%``A Note on accelerating cosmologies from %compactifications and S branes,''
JHEP {\bf 0305}, 028 (2003)
%doi:10.1088/1126-6708/2003/05/028
[hep-th/0304124].

\bibitem{Townsend:2003fx} 
P.~K.~Townsend and M.~N.~R.~Wohlfarth,
%``Accelerating cosmologies from compactification,''
Phys.\ Rev.\ Lett.\  {\bf 91}, 061302 (2003)
%doi:10.1103/PhysRevLett.91.061302
[hep-th/0303097].

\bibitem{Ohta:2003pu} 
N.~Ohta,
%``Accelerating cosmologies from S-branes,''
Phys.\ Rev.\ Lett.\  {\bf 91}, 061303 (2003)
%doi:10.1103/PhysRevLett.91.061303
[hep-th/0303238].

\bibitem{Wohlfarth:2003ni} 
M.~N.~R.~Wohlfarth,
%``Accelerating cosmologies and a phase transition in M %theory,''
Phys.\ Lett.\ B {\bf 563}, 1 (2003)
%doi:10.1016/S0370-2693(03)00609-9
[hep-th/0304089].

\bibitem{Roy:2003nd} 
S.~Roy,
%``Accelerating cosmologies from M / string 
%theory compactifications,''
Phys.\ Lett.\ B {\bf 567}, 322 (2003)
%doi:10.1016/j.physletb.2003.06.060
[hep-th/0304084].

\bibitem{swamp1} 
G.~Obied, H.~Ooguri, L.~Spodyneiko and C.~Vafa,
%``De Sitter Space and the Swampland,''
arXiv:1806.08362 [hep-th].

\bibitem{swamp2} 
P.~Agrawal, G.~Obied, P.~J.~Steinhardt and C.~Vafa,
%``On the Cosmological Implications of the 
%String Swampland,''
Phys.\ Lett.\ B {\bf 784}, 271 (2018)
%doi:10.1016/j.physletb.2018.07.040
[arXiv:1806.09718 [hep-th]].

\bibitem{string} 
M.~B.~Green, J.~H.~Schwarz, and E.~Witten, 
``Superstring theory'', 
Cambridge University Press, Cambridge (1987).

\bibitem{Almeida:2018fwe} 
J.~P.~Beltr\'an Almeida, A.~Guarnizo and 
C.~A.~Valenzuela-Toledo,
%``Arbitrarily coupled $p-$forms 
%in cosmological backgrounds,''
arXiv:1810.05301 [astro-ph.CO].

\bibitem{Ratra1991bn} 
B.~Ratra,
%``Cosmological 'seed' magnetic field from inflation,''
Astrophys.\ J.\  {\bf 391}, L1 (1992).
%doi:10.1086/186384

\bibitem{Bamba2003av} 
K.~Bamba and J.~Yokoyama,
%``Large scale magnetic fields from inflation in dilaton electromagnetism,''
Phys.\ Rev.\ D {\bf 69}, 043507 (2004)
%doi:10.1103/PhysRevD.69.043507
[astro-ph/0310824].
  
\bibitem{Martin2007ue} 
J.~Martin and J.~Yokoyama,
%``Generation of Large-Scale Magnetic Fields in Single-Field Inflation,''
JCAP {\bf 0801}, 025 (2008)
%doi:10.1088/1475-7516/2008/01/025
[arXiv:0711.4307 [astro-ph]].

\bibitem{Yokoyama2008xw} 
S.~Yokoyama and J.~Soda,
%``Primordial statistical anisotropy generated 
%at the end of inflation,''
JCAP {\bf 0808}, 005 (2008)
%doi:10.1088/1475-7516/2008/08/005
[arXiv:0805.4265 [astro-ph]].

\bibitem{Dimopoulos2009am} 
K.~Dimopoulos, M.~Karciauskas and J.~M.~Wagstaff,
%``Vector Curvaton with varying Kinetic Function,''
Phys.\ Rev.\ D {\bf 81}, 023522 (2010)
%doi:10.1103/PhysRevD.81.023522
[arXiv:0907.1838 [hep-ph]].

\bibitem{Dimopoulos2009vu} 
K.~Dimopoulos, M.~Karciauskas and J.~M.~Wagstaff,
%``Vector Curvaton without Instabilities,''
Phys.\ Lett.\ B {\bf 683}, 298 (2010)
%doi:10.1016/j.physletb.2009.12.024
[arXiv:0909.0475 [hep-ph]].

\bibitem{Fujita:2018zbr} 
T.~Fujita, I.~Obata, T.~Tanaka and S.~Yokoyama,
%``Statistically Anisotropic Tensor Modes from Inflation,''
JCAP {\bf 1807}, 023 (2018)
%doi:10.1088/1475-7516/2018/07/023
[arXiv:1801.02778 [astro-ph.CO]].

\bibitem{Watanabe} 
M.~a.~Watanabe, S.~Kanno and J.~Soda,
%``Inflationary Universe with Anisotropic Hair,''
Phys.\ Rev.\ Lett.\  {\bf 102}, 191302 (2009)
%doi:10.1103/PhysRevLett.102.191302
[arXiv:0902.2833 [hep-th]].

\bibitem{Watanabe2} 
M.~a.~Watanabe, S.~Kanno and J.~Soda,
%``Imprints of Anisotropic Inflation on the Cosmic 
%Microwave Background,''
Mon.\ Not.\ Roy.\ Astron.\ Soc.\  {\bf 412}, L83 (2011)
%doi:10.1111/j.1745-3933.2011.01010.x
[arXiv:1011.3604 [astro-ph.CO]].
  
\bibitem{Kanno:2010nr} 
S.~Kanno, J.~Soda and M.~a.~Watanabe,
%``Anisotropic Power-law Inflation,''
JCAP {\bf 1012}, 024 (2010)
%doi:10.1088/1475-7516/2010/12/024
[arXiv:1010.5307 [hep-th]].
    
\bibitem{Ohashi:2013pca} 
J.~Ohashi, J.~Soda and S.~Tsujikawa,
%``Anisotropic power-law k-inflation,''
Phys.\ Rev.\ D {\bf 88}, 103517 (2013)
%doi:10.1103/PhysRevD.88.103517
[arXiv:1310.3053 [hep-th]].  

\bibitem{Gumrukcuoglu2010yc} 
A.~E.~Gumrukcuoglu, B.~Himmetoglu and M.~Peloso,
%``Scalar-Scalar, Scalar-Tensor, and 
%Tensor-Tensor Correlators from Anisotropic Inflation,''
Phys.\ Rev.\ D {\bf 81}, 063528 (2010)
%doi:10.1103/PhysRevD.81.063528
[arXiv:1001.4088 [astro-ph.CO]].

\bibitem{Gum} 
A.~E.~Gumrukcuoglu, C.~R.~Contaldi and M.~Peloso,
%``Inflationary perturbations in anisotropic backgrounds and their imprint on the CMB,''
JCAP {\bf 0711}, 005 (2007)
%doi:10.1088/1475-7516/2007/11/005
[arXiv:0707.4179 [astro-ph]].
   
\bibitem{Himmetoglu2009mk} 
B.~Himmetoglu,
%``Spectrum of Perturbations in Anisotropic Inflationary Universe 
%with Vector Hair,''
JCAP {\bf 1003}, 023 (2010)
%doi:10.1088/1475-7516/2010/03/023
[arXiv:0910.3235 [astro-ph.CO]].

\bibitem{Bartolo} 
N.~Bartolo, S.~Matarrese, M.~Peloso and A.~Ricciardone,
%``Anisotropic power spectrum and bispectrum in the $f(\phi)F^2$ mechanism,''
Phys.\ Rev.\ D {\bf 87}, 023504 (2013)
%doi:10.1103/PhysRevD.87.023504
[arXiv:1210.3257 [astro-ph.CO]].

\bibitem{Namba2012gg} 
R.~Namba,
%``Curvature Perturbations from a Massive Vector Curvaton,''
Phys.\ Rev.\ D {\bf 86}, 083518 (2012)
%doi:10.1103/PhysRevD.86.083518
[arXiv:1207.5547 [astro-ph.CO]].
    
\bibitem{Shiraishi2013vja} 
M.~Shiraishi, E.~Komatsu, M.~Peloso and N.~Barnaby,
%``Signatures of anisotropic sources in the squeezed-limit bispectrum of the cosmic microwave background,''
JCAP {\bf 1305}, 002 (2013)
%doi:10.1088/1475-7516/2013/05/002
[arXiv:1302.3056 [astro-ph.CO]].
 
\bibitem{Fujita2013pgp} 
T.~Fujita and S.~Yokoyama,
%``Higher order statistics of curvature perturbations 
%in IFF model and its Planck constraints,''
JCAP {\bf 1309}, 009 (2013)
%doi:10.1088/1475-7516/2013/09/009
[arXiv:1306.2992 [astro-ph.CO]].

\bibitem{Ohashi2} 
J.~Ohashi, J.~Soda and S.~Tsujikawa,
%``Anisotropic Non-Gaussianity from a Two-Form Field,''
Phys.\ Rev.\ D {\bf 87}, 083520 (2013)
%doi:10.1103/PhysRevD.87.083520
[arXiv:1303.7340 [astro-ph.CO]].

\bibitem{Ito} 
A.~Ito and J.~Soda,
%``Designing Anisotropic Inflation with Form Fields,''
Phys.\ Rev.\ D {\bf 92}, 123533 (2015)
%doi:10.1103/PhysRevD.92.123533
[arXiv:1506.02450 [hep-th]].
 %%CITATION = doi:10.1103/PhysRe

\bibitem{Obata:2018ilf} 
I.~Obata and T.~Fujita,
%``Footprint of Two-Form Field: Statistical Anisotropy in %Primordial Gravitational Waves,''
Phys.\ Rev.\ D {\bf 99}, 023513 (2019)
%doi:10.1103/PhysRevD.99.023513
[arXiv:1808.00548 [astro-ph.CO]].

\bibitem{Almeida:2019xzt} 
J.~P.~B.~Almeida, A.~Guarnizo, R.~Kase, S.~Tsujikawa and C.~A.~Valenzuela-Toledo,
%``Anisotropic inflation with coupled 
%$\boldsymbol{p}$-forms,''
arXiv:1901.06097 [gr-qc].

\bibitem{Ohashi:2013qba} 
J.~Ohashi, J.~Soda and S.~Tsujikawa,
%``Observational signatures of anisotropic 
%inflationary models,''
JCAP {\bf 1312}, 009 (2013)
%doi:10.1088/1475-7516/2013/12/009
[arXiv:1308.4488 [astro-ph.CO]].

\bibitem{Thorsrud:2012mu} 
M.~Thorsrud, D.~F.~Mota and S.~Hervik,
%``Cosmology of a Scalar Field Coupled to Matter and an %Isotropy-Violating Maxwell Field,''
JHEP {\bf 1210}, 066 (2012)
%doi:10.1007/JHEP10(2012)066
[arXiv:1205.6261 [hep-th]].

\bibitem{Koivisto:2007bp} 
T.~Koivisto and D.~F.~Mota,
%``Accelerating Cosmologies with an Anisotropic 
%Equation of State,''
 Astrophys.\ J.\  {\bf 679}, 1 (2008)
%doi:10.1086/587451
[arXiv:0707.0279 [astro-ph]].

\bibitem{Koivisto:2008ig} 
T.~Koivisto and D.~F.~Mota,
%``Anisotropic Dark Energy: Dynamics of Background 
%and Perturbations,''
JCAP {\bf 0806}, 018 (2008)
%doi:10.1088/1475-7516/2008/06/018
[arXiv:0801.3676 [astro-ph]].

\bibitem{Battye:2009ze} 
R.~Battye and A.~Moss,
%``Anisotropic dark energy and CMB anomalies,''
Phys.\ Rev.\ D {\bf 80}, 023531 (2009)
%doi:10.1103/PhysRevD.80.023531
[arXiv:0905.3403 [astro-ph.CO]].

\bibitem{Appleby:2009za} 
S.~Appleby, R.~Battye and A.~Moss,
%``Constraints on the anisotropy of dark energy,''
Phys.\ Rev.\ D {\bf 81}, 081301 (2010)
%doi:10.1103/PhysRevD.81.081301
[arXiv:0912.0397 [astro-ph.CO]].

\bibitem{Campanelli:2010zx} 
L.~Campanelli, P.~Cea, G.~L.~Fogli and A.~Marrone,
%``Testing the Isotropy of the Universe with Type 
%Ia Supernovae,''
Phys.\ Rev.\ D {\bf 83}, 103503 (2011)
%doi:10.1103/PhysRevD.83.103503
[arXiv:1012.5596 [astro-ph.CO]].

\bibitem{Appleby:2012as} 
S.~A.~Appleby and E.~V.~Linder,
%``Probing dark energy anisotropy,''
Phys.\ Rev.\ D {\bf 87}, 023532 (2013)
%doi:10.1103/PhysRevD.87.023532
[arXiv:1210.8221 [astro-ph.CO]].

\bibitem{Betoule:2014frx} 
M.~Betoule {\it et al.} [SDSS Collaboration],
%``Improved cosmological constraints from a joint 
%analysis of the SDSS-II and SNLS supernova samples,''
Astron.\ Astrophys.\  {\bf 568}, A22 (2014)
%doi:10.1051/0004-6361/201423413
[arXiv:1401.4064 [astro-ph.CO]].

\bibitem{Aghanim:2018eyx} 
N.~Aghanim {\it et al.} [Planck Collaboration],
%``Planck 2018 results. VI. Cosmological parameters,''
arXiv:1807.06209 [astro-ph.CO].

\bibitem{Hu05} 
D.~Giannakis and W.~Hu,
%``Kinetic unified dark matter,''
Phys.\ Rev.\ D {\bf 72}, 063502 (2005)
%doi:10.1103/PhysRevD.72.063502
[astro-ph/0501423].

\bibitem{Arroja} 
F.~Arroja and M.~Sasaki,
%``A note on the equivalence of a barotropic perfect fluid 
%with a K-essence scalar field,''
Phys.\ Rev.\ D {\bf 81}, 107301 (2010)
%doi:10.1103/PhysRevD.81.107301
[arXiv:1002.1376 [astro-ph.CO]].

\bibitem{KT14} 
R.~Kase and S.~Tsujikawa,
%``Cosmology in generalized Horndeski theories 
%with second-order equations of motion,''
Phys.\ Rev.\ D {\bf 90}, 044073 (2014)
%doi:10.1103/PhysRevD.90.044073
[arXiv:1407.0794 [hep-th]].

\bibitem{Bean} 
R.~Bean, S.~H.~Hansen and A.~Melchiorri,
%``Early universe constraints on a primordial scaling field,''
Phys.\ Rev.\ D {\bf 64}, 103508 (2001)
%doi:10.1103/PhysRevD.64.103508
[astro-ph/0104162].

\bibitem{Ohashi:2009xw} 
J.~Ohashi and S.~Tsujikawa,
%``Assisted dark energy,''
Phys.\ Rev.\ D {\bf 80}, 103513 (2009)
%doi:10.1103/PhysRevD.80.103513
[arXiv:0909.3924 [gr-qc]].

\bibitem{Ade15} 
P.~A.~R.~Ade {\it et al.} [Planck Collaboration],
%``Planck 2015 results. XIV. Dark energy and modified gravity,''
Astron.\ Astrophys.\  {\bf 594}, A14 (2016)
%doi:10.1051/0004-6361/201525814
[arXiv:1502.01590 [astro-ph.CO]].

\bibitem{Wald} 
R.~M.~Wald,
%``Asymptotic behavior of homogeneous cosmological models 
%in the presence of a positive cosmological constant,''
Phys.\ Rev.\ D {\bf 28}, 2118 (1983).
%doi:10.1103/PhysRevD.28.2118

\bibitem{Starobinsky1982mr} 
A.~A.~Starobinsky,
%``Isotropization of arbitrary cosmological expansion given 
%an effective cosmological constant,''
JETP Lett.\  {\bf 37}, 66 (1983).

\end{thebibliography}
\end{document}